\documentclass[reprint,amsmath,amssymb,
 aps,prl]{revtex4-2}

\usepackage{graphicx}
\usepackage{dcolumn}
\usepackage{bm,ulem}
\usepackage[pdftex,
            colorlinks=true,
            citecolor=blue,
            linkcolor=blue]{hyperref}
\usepackage{appendix}

\begin{document}

\preprint{APS/123-QED}

\title{Superconducting meron phase in locally noncentrosymmetric superconductors}

\author{Akihiro Minamide}
\email{minamide.akihiro.34n@st.kyoto-u.ac.jp}
\affiliation{Department of Physics, Kyoto University, Kyoto 606-8502, Japan}

\author{Youichi Yanase}
\affiliation{Department of Physics, Kyoto University, Kyoto 606-8502, Japan}

\date{\today}

\begin{abstract}
Theory of the superconducting parity transition is extended by incorporating the vortex degree of freedom.
We employ the bilayer Rashba model representing locally noncentrosymmetric layered superconductors and derive the Ginzburg-Landau free energy functional.
This formulation reveals the parity transition, where the even-parity superconducting state changes to the odd-parity one upon increasing the magnetic field under the vortex states.
The $H$-$T$ phase diagram of $\mathrm{CeRh_2As_2}$ is quantitatively reproduced and a novel superconducting state with a meron (half-skyrmion) lattice pseudospin texture is predicted.
\end{abstract}

\maketitle

\textit{Introduction. ---}
Since the discovery of heavy-fermion superconductors~\cite{steglich1979superconductivity} and high-$T_{\rm c}$ superconductors~\cite{bednorz1986possible}, vigorous efforts have been made to extend the BCS theory~\cite{bardeen1957microscopic} and find unconventional superconducting states by incorporating additional degrees of freedom of Cooper pairs.
In recent years, superconductivity in locally noncentrosymmetric (LNC) crystals~\cite{mockli2022unconventional,fischer2023superconductivity} has attracted more attention.
The crystals preserve the global inversion symmetry, but the inversion center does not lie on the atom position.
Then, sublattice degrees of freedom can be assigned to the atomic sites which are interchanged by inversion operation.
The LNC crystal structures are found in many materials, including the one-dimensional zigzag/ladder structure of $\mathrm{UGe_2,\ URhGe,\ UCoGe}$~\cite{aoki2011ferromagnetism} and $\mathrm{UTe_2}$~\cite{aoki2022unconventional}, the two-dimensional honeycomb structure of graphene, $\mathrm{SrPtAs}$~\cite{nishikubo2011superconductivity} and transition metal dichalcogenides~\cite{wilson1969transition}, and the three-dimensional layered structure of high-$T_{\rm c}$ cuprates~\cite{mukuda2011high} and artificial superlattices $\mathrm{CeCoIn_5/YbCoIn_5}$~\cite{mizukami2011extremely}. 
One of the various intriguing phenomena in LNC superconductors is the superconducting parity transition~\cite{yoshida2012pair}.
For layered systems under the out-of-plane magnetic field, a phase diagram with two superconducting phases has been theoretically proposed: the Bardeen-Cooper-Schrieffer (BCS) state, where the superconducting order parameter is sublattice-symmetric, is stable at zero field, while the pair-density-wave (PDW) state, where the order parameter is sublattice-antisymmetric, is energetically favored under a sufficiently high field.
The latter is an odd-parity spin-singlet superconducting state, which is outside the framework of the conventional BCS theory, and a candidate of topological superconductors~\cite{yoshida2015topological,nogaki2021topological,ishizuka2023correlation}.
The parity transition and resulting phenomena have been studied not only in the layered systems~\cite{yoshida2012pair,yoshida2014parity,sigrist2014superconductors,higashi2016robust,skurativska2021spin,nogaki2022even,hackner2023bardasis,lee2023linear,szabo2023superconductivity,nogaki2023field,szabo2024effects} but also in other LNC systems~\cite{fischer2011superconductivity,watanabe2015odd,sumita2016superconductivity,nakamura2017odd}.

In 2021, heavy-fermion superconductivity in $\mathrm{CeRh_2As_2}$ was discovered~\cite{khim2021field}.
The crystal structure represented by the nonsymmorphic centrosymmetric space group $P4/nmm\ (\#129)$
consists of $\mathrm{Ce}$ layers, which locally violate the inversion symmetry because they are sandwiched between two inequivalent $\mathrm{Rh}$-$\mathrm{As}$ blocks.
Quite remarkably, the superconducting state shows the unusual $H$-$T$ phase diagram, in which the phase transition between the low-field and high-field superconducting phases occurs under the $c$-axis magnetic field.
Its resemblance to the theoretical prediction~\cite{yoshida2012pair} suggests realization of the parity transition in $\mathrm{CeRh_2As_2}$, and the subsequent experimental results~\cite{kimura2021optical,onishi2022low,landaeta2022field,mishra2022anisotropic,semeniuk2023decoupling,siddiquee2023pressure,chen2023coexistence,wu2023fermi,pfeiffer2023pressure,pfeiffer2023exposing,christovam2024spectroscopic,chen2024exploring} are compatible with this scenario.
However, there remain several features that are not understood well in the normal and superconducting states of $\mathrm{CeRh_2As_2}$.
Specific heat measurements revealed the presence of a weak anomaly above the superconducting transition temperature~\cite{khim2021field,semeniuk2023decoupling,chajewski2024discovery}, which is considered to be a phase transition to the electric quadrupole-density-wave state~\cite{hafner2022possible}.
In addition, nuclear magnetic and quadrupolar resonance measurements have shown the antiferromagnetic order inside the superconducting phase~\cite{kibune2022observation,kitagawa2022two,ogata2023parity,ogata2023investigation}.
Moreover, the proximity to the quantum critical point has been discussed~\cite{khim2021field,pfeiffer2023pressure}, but its relation to the origin of superconductivity is elusive.

Aside from a deep understanding of the competing and coexisting orders, we revisit the general issue of parity transition.
In general, superconductivity is affected by a magnetic field through the Pauli and orbital depairing effects.
The former arises from the Zeeman splitting, while the latter is due to the formation of quantum vortices.
Almost all the previous theoretical studies on LNC superconductors have focused on the pure Pauli limit, and the effect of the vortex lattice formation has only been treated phenomenologically~\cite{mockli2018orbitally,schertenleib2021unusual,fischer2023superconductivity,khim2021field}.
However, quantum vortices have long been studied for noncentrosymmetric superconductors~\cite{kaur2005helical,oka2006surface,dimitrova2007theory,hiasa2008orbital,iniotakis2008fractional,matsunaga2008modulated,hiasa2009vortex,kashyap2013vortices,nikolic2014vortices,dan2015quasiclassical}, in which the antisymmetric spin-orbit coupling (ASOC) plays an essential role in electronic states as in the LNC superconductors.
Therefore, the study of quantum vortices has great potential for the discovery of interesting features in the LNC superconductors.
In this paper, we deal with the vortex degree of freedom in the LNC superconductors microscopically.
The Ginzburg-Landau (GL) free energy functional is derived from the bilayer Rashba model.
The occurrence of parity transition is confirmed, and the phase diagram of $\mathrm{CeRh_2As_2}$ is quantitatively reproduced.
Furthermore, it is shown that a novel superconducting state characterized by the meron lattice pseudospin texture is stabilized near the multicritical point.
We use the unit $\hbar=c=k_{\rm B}=1$ throughout this paper.

\begin{figure}[htbp]
  \centering  \includegraphics[width=0.30\textwidth]{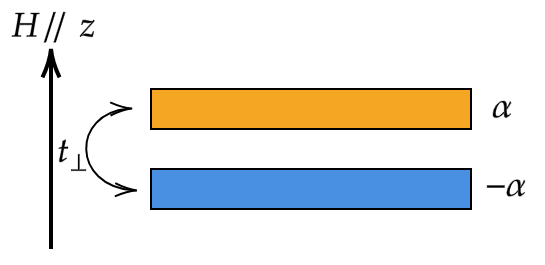}
  \caption{Schematic figure of LNC bilayer systems. Interlayer hopping $t_\perp$, external magnetic field $H$, and layer-dependent ASOC $\alpha$ are illustrated.
  }
\label{fig:bilayer_system_schematic}
\end{figure}

\textit{Model and method. ---}
We start from the bilayer Rashba model, which describes LNC layered systems depicted schematically in Fig.~\ref{fig:bilayer_system_schematic}:
\begin{align}
    \label{eq:mo-1}
    &\mathcal{H}
    =
    \mathcal{H}_{0}
    +\mathcal{H}_{\mathrm{int}},\\
    \label{eq:mo-2}
    &\mathcal{H}_{0}
    =
    \sum_{\bm{k},s,m}
    \xi(\bm{k})c^{\dagger}_{\bm{k}sm}c_{\bm{k}sm}
    +
    t_{\perp}
    \sum_{\bm{k},s}
    \left(c_{\bm{k}s1}^{\dagger}c_{\bm{k}s2}
    +\mathrm{H.c.}\right) \notag\\
    &+
    \sum_{\bm{k},s, s',m}
    \alpha_m \bm{g}(\bm{k})\cdot\bm{\sigma}_{ss'}
    c^{\dagger}_{\bm{k}sm}
    c_{\bm{k}s'm}
    -\mu_{\rm B} H
    \sum_{\bm{k},s,m}
    sc_{\bm{k}sm}^{\dagger}c_{\bm{k}sm},\\
    \label{eq:mo-3}
    &\mathcal{H}_{\mathrm{int}}
    =
    -V
    \sum_{\bm{k},\bm{k'},\bm{q},m}
    c_{\bm{k}_{+}\uparrow m}^{\dagger}
    c_{-\bm{k}_{-}\downarrow m}^{\dagger}
    c_{-\bm{k'}_{-}\downarrow m}
    c_{\bm{k'}_{+}\uparrow m}.
\end{align}
Here, $c_{\bm{k}sm} (c_{\bm{k}sm}^{\dagger})$ is an annihilation (creation) operator for an electron with two-dimensional momentum $\bm{k}$ and spin $s (=\uparrow,\downarrow)$ on the $m (=1,2)$th layer.
$\bm{\sigma}=(\sigma_x,\sigma_y,\sigma_z)$ are the Pauli matrices, and $\bm{k}_{\pm}$ implies $\bm{k}\pm \bm{q}/2$.
The isotropic in-plane energy dispersion $\xi(\bm{k})=k^2/(2m)-\mu$ is assumed, where $m$ is the electron mass and $\mu$ is the chemical potential.
Originating from the local lack of inversion symmetry of each layer, the ASOC arises with the Rashba-type $g$ vector $\bm{g}(\bm{k})=\bm{k}\times \hat{z}/k_{\rm F}$.
The coupling constants $\alpha_m$ are layer-dependent to preserve the global inversion symmetry and given as $(\alpha_1,\alpha_2)=(\alpha,-\alpha)$ for our bilayer model.
It is useful to define a parameter $\chi$~\cite{lee2023linear} by $e^{i\chi}=(t_\perp+i\alpha)/\sqrt{t_\perp^2+\alpha^2}$ to represent the relative strength of ASOC and interlayer hopping $t_\perp$.
The magnetic field is applied parallel to the $z$-axis, and $\bm{A}(\bm{r})=Hx\hat{\bm{y}}$ is the vector potential in the Landau gauge, where $\bm{r}=(x,y)$ is a two-dimensional spatial coordinate on the layers.
The orbital depairing effect can be incorporated through the substitution $\bm{k}\to \bm{k}+e\bm{A}$, where $e>0$ is the elementary charge.
We focus on the type-II limit and ignore the screening by the supercurrent. 
For simplicity, the $s$-wave attractive interaction is assumed.
Following previous studies~\cite{khim2021field,cavanagh2022nonsymmorphic,lee2023linear}, we impose the condition for energy scales
\begin{equation}
    \label{eq:mo-4}
    \mu_{\rm B} H,T_{{\rm c}0}\ll \alpha,t_\perp\ll E_{\rm F},
\end{equation}
where $T_{{\rm c}0}$ is the transition temperature at zero magnetic field and $E_{\rm F}$ is the Fermi energy.

Implementing the Hubbard-Stratonovich transform~\cite{hubbard1959calculation}, we introduce the auxiliary complex field $\Delta$.
Since $\Delta$ is small near the transition line, the thermodynamic potential can be expanded in powers of $\Delta$.
Following the similar procedure described in Refs.~\cite{dimitrova2007theory,adachi2003effects,hiasa2009vortex,hiasa2008orbital,adachi2015possible}, the GL free energy functional is constructed as 
\begin{equation}
    \label{eq:mo-12}
    \mathcal{F}[\Delta,\bm{A}]=\mathcal{F}^{(2)}+\mathcal{F}^{(4)}+\mathcal{O}(|\Delta|^6),
\end{equation}
where $\mathcal{F}^{(n)}$ is the $n$th order term of $\Delta$~\cite{supplemental}.

The order parameter on each layer $\Delta_m (m=1,2)$ is decomposed into the sublattice-symmetric component $\Delta_e$ and sublattice-antisymmetric component $\Delta_o$.
The quadratic term of the GL free energy functional is 
\begin{equation}
    \label{eq:mo-13}
    \mathcal{F}^{(2)}
    =
    \sum_{j=e,o}\int d^2\bm{r}
    \ \Delta^*_j(\bm{r})\left(\frac{2}{V}-\hat{\mathcal{K}}_{j}^{(2)}(\bm{\Pi})\right)\Delta_j(\bm{r}),
\end{equation}
where $\bm{\Pi}=-i\nabla+2e\bm{A}$ is the gauge-invariant operator for Cooper pairs, and differential operators $\hat{\mathcal{K}}_{j}^{(2)}(\bm{\Pi})$ act on $\Delta_j$.
Interband pairings are negligible since they are unstable under the condition Eq.~\eqref{eq:mo-4}.
The magnitudes of the attractive interaction are allowed to be different for $\Delta_e$- and $\Delta_o$-channels depending on the origin of superconductivity, e.g., multipole fluctuations~\cite{nogaki2023field}.
Therefore, we replace $V$ in Eq.\eqref{eq:mo-13} with $V_e$ or $V_o$.
In the presence of a magnetic field, the eigenfunctions of $\hat{\mathcal{K}}_{j}^{(2)}(\bm{\Pi})$ are basis functions of Landau levels (LLs).
Let 
\begin{equation}
    \label{eq:mo-16}
    \psi_{Nq}(\bm{r})
    =\frac{H_N(x/r_H+qr_H)}{\sqrt{2^NN!}}
    \frac{e^{-(x/r_H+qr_H)^2/2+iqy}}{\sqrt{\pi^{1/2}r_HL_y}},
\end{equation}
be the $N$th LL basis function corresponding to the Landau gauge with momentum $q$ in the $y$-direction, where $r_H=(2eH)^{-1/2}$ is the magnetic length and $H_N$ is the $N$th Hermite polynomial.
$\Delta_j$ is expanded as $\Delta_j(\bm{r})=\sum_{N,q}\Delta_j(N,q)\psi_{Nq}(\bm{r})$.
Then, we reach the expression of $\mathcal{F}^{(2)}$,
\begin{align}
    \label{eq:mo-22}
    &\mathcal{F}^{(2)}
    =2N(0)
    \sum_{N,q,j}
    E_{N}^{\,j}
    |\Delta_j(N,q)|^2,\\
    &E_{N}^{\,e}
    =
    \ln\left(\frac{T}{T_{\mathrm{c}0}^{e}}\right)+\int^{\infty}_{0}\frac{d\eta}{\sinh(2\eta)}
    \tilde{\mathcal{K}}^{(2)}_{e}(\eta,N),\\
    &E_{N}^{\,o}
    =
    \sin^2\chi\left[\ln\left(\frac{T}{T_{\mathrm{c}0}^{o}}\right)+\int^{\infty}_{0}\frac{d\eta}{\sinh(2\eta)}
    \tilde{\mathcal{K}}^{(2)}_{o}(\eta,N)\right],\\
    \label{eq:mo-23}
    &\tilde{\mathcal{K}}^{(2)}_{e}(\eta,N)
    =1-\cos(2h\eta\cos\chi)
    e^{-\eta^2/4\zeta_H^2}
    L_N\left(\frac{\eta^2}{2\zeta_H^2}\right),\\
    \label{eq:mo-24}
    &\tilde{\mathcal{K}}^{(2)}_{o}(\eta,N)
    =1-
    e^{-\eta^2/4\zeta_H^2}
    L_N\left(\frac{\eta^2}{2\zeta_H^2}\right),
\end{align}
where $v_{\rm F}$ is the Fermi velocity, $N(0)$ is the density of states on the Fermi level per unit area, $T_{\mathrm{c}0}^{j}$ is the zero-field transition temperature, $\zeta_H=\pi Tr_H/v_{\rm F}$, $h=\mu_{\rm B} H/\pi T$, and $L_N$ is the $N$th Laguerre polynomial.
Unlike noncentrosymmetric superconductors~\cite{hiasa2009vortex}, even when the magnetic field is perpendicular to layers, the Pauli paramagnetism is effective on the $\Delta_e$ pairing due to the interlayer coupling $t_\perp$~\cite{maruyama2012locally}, while in the $\Delta_o$ pairing state, the paramagnetic depairing effect is completely suppressed.
The Maki parameter~\cite{maki1966effect} is given by $\alpha_M=2\sqrt{2}\pi T_{\mathrm{c}0}^{e}\cos\chi/(mv_{\rm F}^2)$ for our model.

Next, the quartic term of the GL free energy functional takes the form
\begin{align}
    \label{eq:mo-27}
    \mathcal{F}^{(4)}
    =&\frac{1}{2}
    \sum_{\{j_i\}}\int d^2\bm{r}\ 
    \mathrm{Re}
    \left[\hat{\mathcal{K}}^{(4)}(\{\bm{\Pi}_i\},\{j_i\})
    \right.\notag\\
    &\ \left.\times\Delta_{j_1}^*(\bm{r}_{1})\Delta_{j_2}(\bm{r}_{2})\left.\Delta_{j_3}^*(\bm{r}_{3})
    \Delta_{j_4}(\bm{r}_{4})\right|_{\bm{r}_{i}=\bm{r}}\right].
\end{align}
The general expression that includes higher-order LLs is complicated.
Hereafter, we assume that the pair potential $\Delta$ consists only of the lowest LL ($N=0$).
This assumption is reasonable near the upper critical field when the paramagnetic depairing effect is not too strong.

\textit{Phase diagram. ---}
Let us show the phase diagram revealing the parity transition.
Assuming the continuous superconducting transition, we evaluate the upper critical field $H_{\rm c2}^{j,N}(T)$ for even-parity ($j=e$) and odd-parity ($j=o$) pairing states in the $N$th LL as the magnetic field where $E_{N}^{\,j}$ alters its sign from positive to negative.

Before showing our results, we discuss previous theories on the parity transition.  
The first study~\cite{yoshida2012pair} assumed spatially uniform order parameters and concluded that the phase diagram includes two superconducting phases under the out-of-plane magnetic field.
One is the BCS state $(\Delta_1,\Delta_2)=(\Delta,\Delta)\Leftrightarrow(\Delta_e,\Delta_o)=(\Delta,0)$, and the other is the PDW state $(\Delta_1,\Delta_2)=(\Delta,-\Delta)\Leftrightarrow(\Delta_e,\Delta_o)=(0,\Delta)$. 
The PDW state is unique to the LNC superconductors, where odd-parity spin-singlet pairing state beyond the standard BCS theory is allowed by the sublattice degree of freedom.
Since the BCS state is stabilized by the interlayer Josephson coupling at zero magnetic field, we set $T_{\mathrm{c}0}^{o}/T_{\mathrm{c}0}^{e}=0.9 (<1)$, consistent with CeRh$_2$As$_2$~\cite{khim2021field}.

\begin{figure}[htbp]
  \centering
  \includegraphics[width=0.48\textwidth]{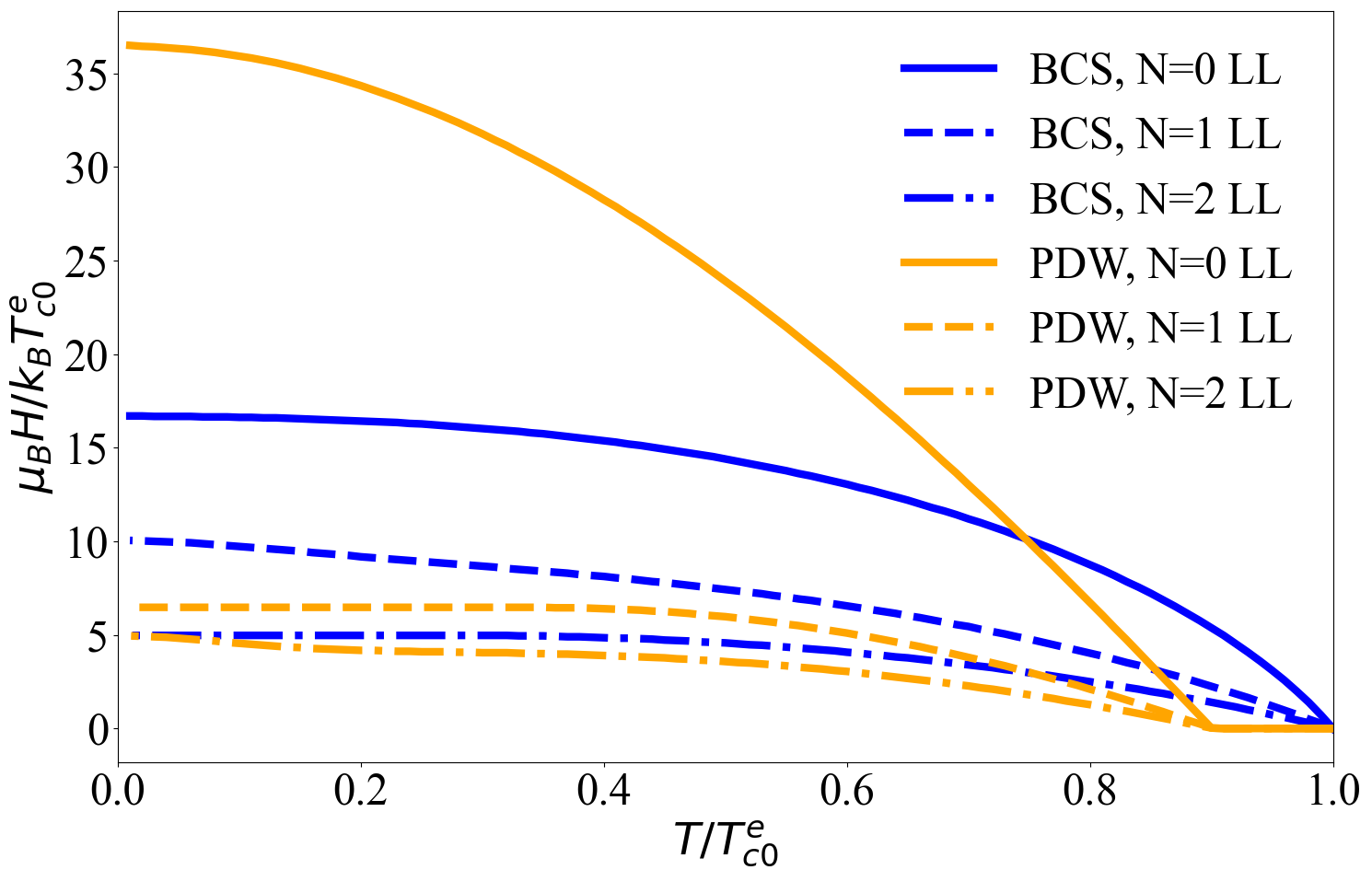}
  \caption{Upper critical fields of the BCS (blue) and PDW (orange) states for $\alpha/t_\perp=15.0$ and $\alpha_M=4.8$.
  The solid, dashed, and dot-dashed lines represent the $N=0,1,2$ LLs, respectively.
  }
\label{fig:phase_diagram_1}
\end{figure}

Analyses based on phenomenological GL free energy~\cite{mockli2018orbitally,schertenleib2021unusual} took into account magnetic flux penetration and verified the parity transition for a sufficiently large Maki parameter, but temperature- and magnetic-field-dependence of some GL coefficients were neglected, making a quantitative investigation difficult.
Our microscopic formulation incorporates comprehensive $T$- and $H$-dependence without ambiguity and allows us to calculate the upper critical fields of spatially inhomogeneous vortex states,
which are shown in Fig.~\ref{fig:phase_diagram_1} for the BCS and PDW phases up to the third lowest LL.
Since the Maki parameter $\alpha_M=4.8$ is not so large, the upper critical fields of higher LLs lie well below that of the lowest LL.
Therefore, as mentioned earlier, it is justified to ignore higher LLs.
Focusing on the lowest LLs, we notice that the two upper critical fields cross, and the BCS (PDW) state is stable in the low-field (high-field) region.
This reveals the parity transition which can be understood from the fact that the PDW state is not subject to the paramagnetic depairing {\it i.e.} it is more robust against the magnetic field~\cite{maruyama2012locally}.
Furthermore, by choosing the parameters $\alpha/t_\perp=15.0$, $\alpha_M=4.8$ as in Fig.~\ref{fig:phase_diagram_1}, our calculations place the multicritical point at $(T/T_{\mathrm{c}0}^{e},\mu_{\rm B} H/k_{\rm B}T_{\mathrm{c}0}^{e})\simeq (0.74,10.1)$ and the upper critical field of the PDW state in the zero-temperature limit at $\mu_{\rm B} H/k_{\rm B}T_{\mathrm{c}0}^{e}\simeq 36.5$, both of which are in good agreement with the experimental result for $\mathrm{CeRh_2As_2}$~\cite{khim2021field}.
The large ratio $\alpha/t_\perp$ compared to previous studies~\cite{yoshida2012pair,nogaki2022even} may be reasonable, if the interlayer hopping is suppressed at the Brillouin zone boundary due to the nonsymmorphic crystal structure~\cite{nica2015glide,cavanagh2022nonsymmorphic,sumita2016superconductivity}. The Fermi surfaces of $\mathrm{CeRh_2As_2}$ were reported to exist around the boundary~\cite{hafner2022possible,ishizuka2023correlation,chen2023coexistence,wu2023fermi}.
Also, the Maki parameter $\alpha_M=4.8$ is comparable to the experimentally estimated value for $\mathrm{CeRh_2As_2}$~\cite{kitagawa2022two} and other heavy-fermion superconductors~\cite{matsuda2007fulde}.
There is no upturn of $H_{\rm c2}^{o,0}(T)$ in the low-temperature region, which agrees well with the experimental result~\cite{khim2021field}, in contrast to the previous studies in the Pauli limit~\cite{yoshida2012pair}. This is because the PDW state is suppressed by the moderate orbital depairing effect~\cite{khim2021field}.

\begin{figure}[htbp]
    \centering
    \includegraphics[width=0.48\textwidth]{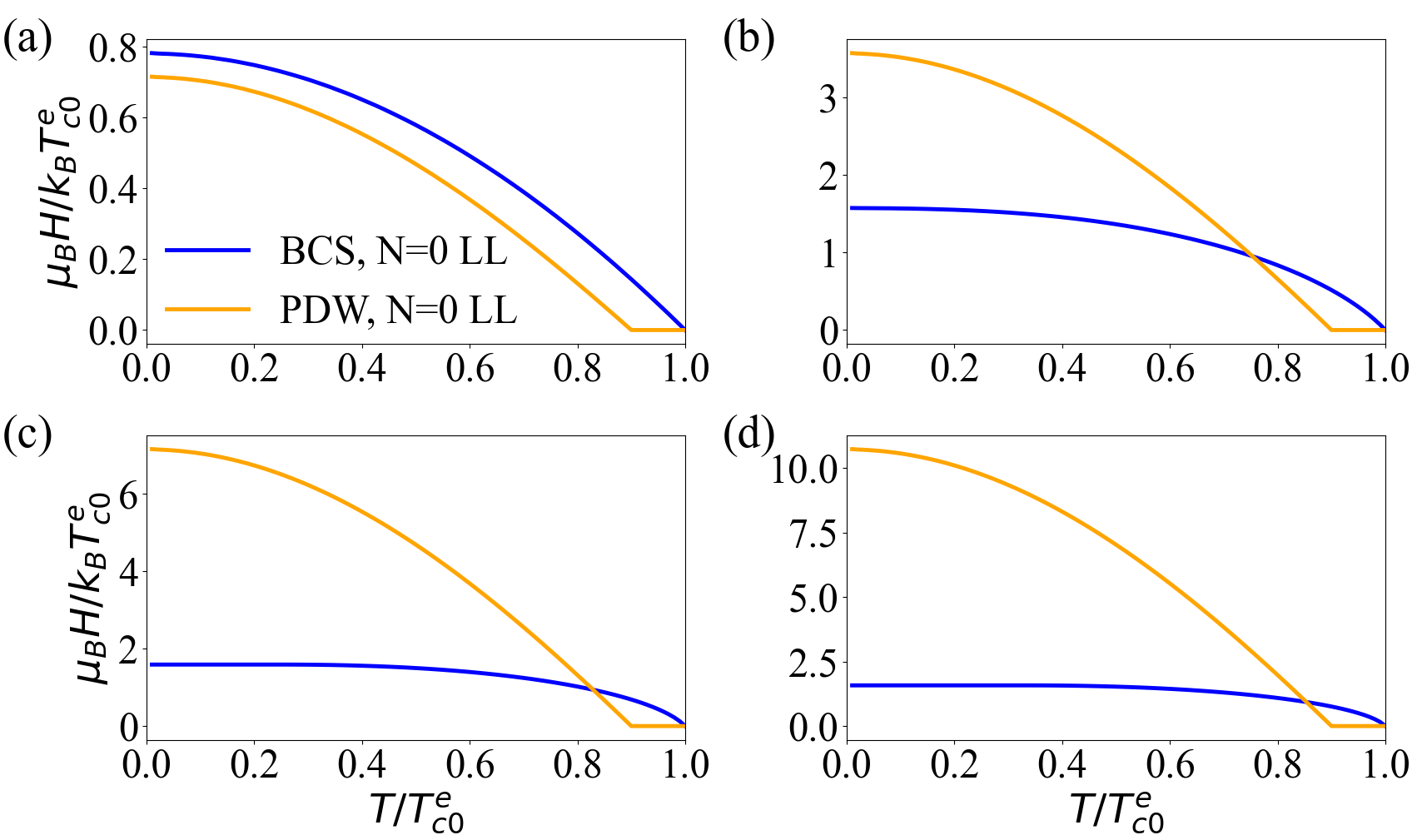}
    \caption{Upper critical fields of the BCS (blue) and PDW (orange) states in the lowest LL for (a) $\alpha_M=1.0$, (b) $\alpha_M=5.0$, (c) $\alpha_M=10.0$, and (d) $\alpha_M=15.0$. We fix $\alpha/t_\perp=1.0$.
  }
    \label{fig:phase_diagram_2}
\end{figure}

The phase diagram is not sensitive to the parameter $\alpha/t_\perp$.
This does not contradict the previous studies~\cite{yoshida2012pair,nogaki2022even}, in which the PDW phase is broadened with increasing $\alpha/t_\perp$. This is because the ratio $T_{\mathrm{c}0}^{o}/T_{\mathrm{c}0}^{e}$ is an independent parameter and set constant in our formulation.
On the other hand, the Maki parameter $\alpha_M$ essentially changes the phase diagram (Fig.~\ref{fig:phase_diagram_2}).  
As increasing $\alpha_M$, the PDW phase becomes stable in the wider region of the phase diagram.
The parity transition occurs for the moderate Maki parameter $\alpha_M\gtrsim 1.4$.

\textit{Superconducting meron phase. ---}
Finally, we study the vortex states realized inside the superconducting phase. 
The superconducting condensation energy in equilibrium is obtained as
\begin{equation}
    \label{eq:sk-1}
    \mathcal{F}_c=-\frac{(\mathcal{F}^{(2)})^2}{2\mathcal{F}^{(4)}},
\end{equation}
when $\mathcal{F}^{(2)}<0 $ and $ \mathcal{F}^{(4)}>0$ are satisfied.
The equilibrium vortex lattice structure is determined by minimizing $\mathcal{F}_c$.
Ideally, we need to solve this problem with respect to all the vortex lattice structures.
However, in order to reduce the computational cost, a triangular lattice is assumed because it is generally most favored in $s$-wave superconductors.

We especially focus on the region near the multicritical point, where both even- and odd-parity order parameters may be active. The order parameters of the BCS and PDW states are sublattice-symmetric or sublattice-antisymmetric.
Here, in addition to the triangular vortex lattices of these states, we study the stability of a third state, where the sublattice-symmetric and -antisymmetric components coexist and space inversion symmetry is spontaneously broken.
This state is hereafter referred to as the superconducting meron (SCM) state and is defined with the order parameter
\begin{align}
    \label{eq:sk-2}
    \Delta_m(\bm{r})
    =&\frac{1}{\sqrt{\pi^{1/2}r_HL_y}}
    \sum_{n=-\infty}^{\infty}
    \left[e^{-(x/r_H+q_{2n}r_H)^2/2+iq_{2n}y}\right.\notag\\
    &\left.+(-1)^{m+1}ie^{-(x/r_H+q_{2n+1}r_H)^2/2+iq_{2n+1}y}
    \right],
\end{align}
where $q_n=2\pi n/L_y$.
The positions of vortex cores on the two layers $\bm{r}_m$ are given by $\bm{r}_1=m_1\bm{a}_1+m_2\bm{a}_2$ and $\bm{r}_2=n_1\bm{a}_1+(n_2+1/2)\bm{a}_2$, where $m_1,\ m_2,\ n_1,\ n_2$ are integers, $\bm{a}_1=(\sqrt{3}/2,1/2)a$ and $\bm{a}_2=(0,1)a$ are the primitive translation vectors of the triangular vortex lattice, and $a=2\sqrt{\pi}3^{-1/4}r_H$ is the vortex lattice constant.
That is, the vortex cores are shifted between two layers by half of $\bm{a}_2$.

\begin{figure}[htbp]

    \centering
    \includegraphics[width=0.48\textwidth]{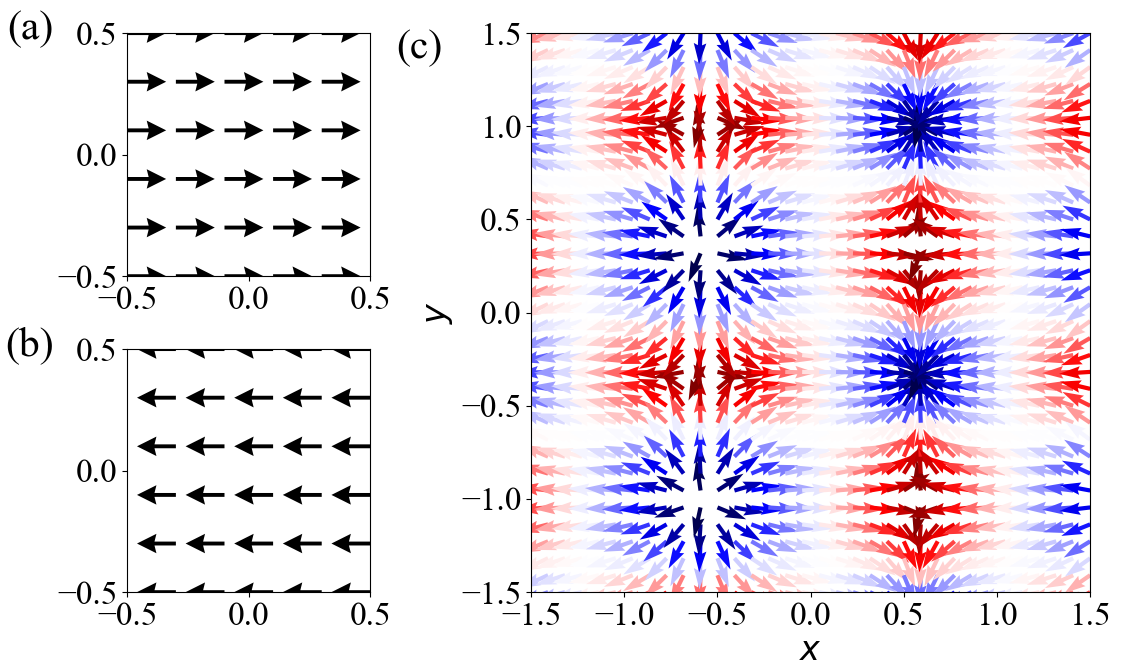}
    \caption{The pseudospin texture of (a) the BCS, (b) PDW, and (c) SCM states.
    The arrows show $(S_x, S_y)/\sqrt{S_x^2+S_y^2}$.
    In (c), $S_z$ is represented by color gradation from red ($S_z=1$) to blue ($S_z=-1$).
  }
    \label{fig:pseudo-spin-texture}

\end{figure}

To characterize the superconducting states discussed above, the pseudospin density is introduced based on the two-component order parameters by $\bm{S}(\bm{r})=\chi^{\dagger}(\bm{r})\bm{\sigma}\chi(\bm{r})$~\cite{kasamatsu2005spin,dimitrova2007theory},
where $\chi=(\chi_1,\chi_2)^{T}$ with $\chi_m=\Delta_m/\sqrt{|\Delta_1|^2+|\Delta_2|^2}$.
In the BCS and PDW states, pseudospins are uniformly oriented in the $x$-direction.
On the other hand, the SCM state has a skyrmion-lattice-like structure, as shown in Fig.~\ref{fig:pseudo-spin-texture}.
Strictly speaking, the merons, with half the topological skyrmion number of skyrmions~\cite{gobel2021beyond}, form a lattice with alternating regions of positive and negative $S_z$. The sublattice degree of freedom extends the manifold of order parameters and allows for topological defects beyond the vortex, in this case superconducting merons.

\begin{figure}[htbp]
  \centering
  \includegraphics[width=0.48\textwidth]{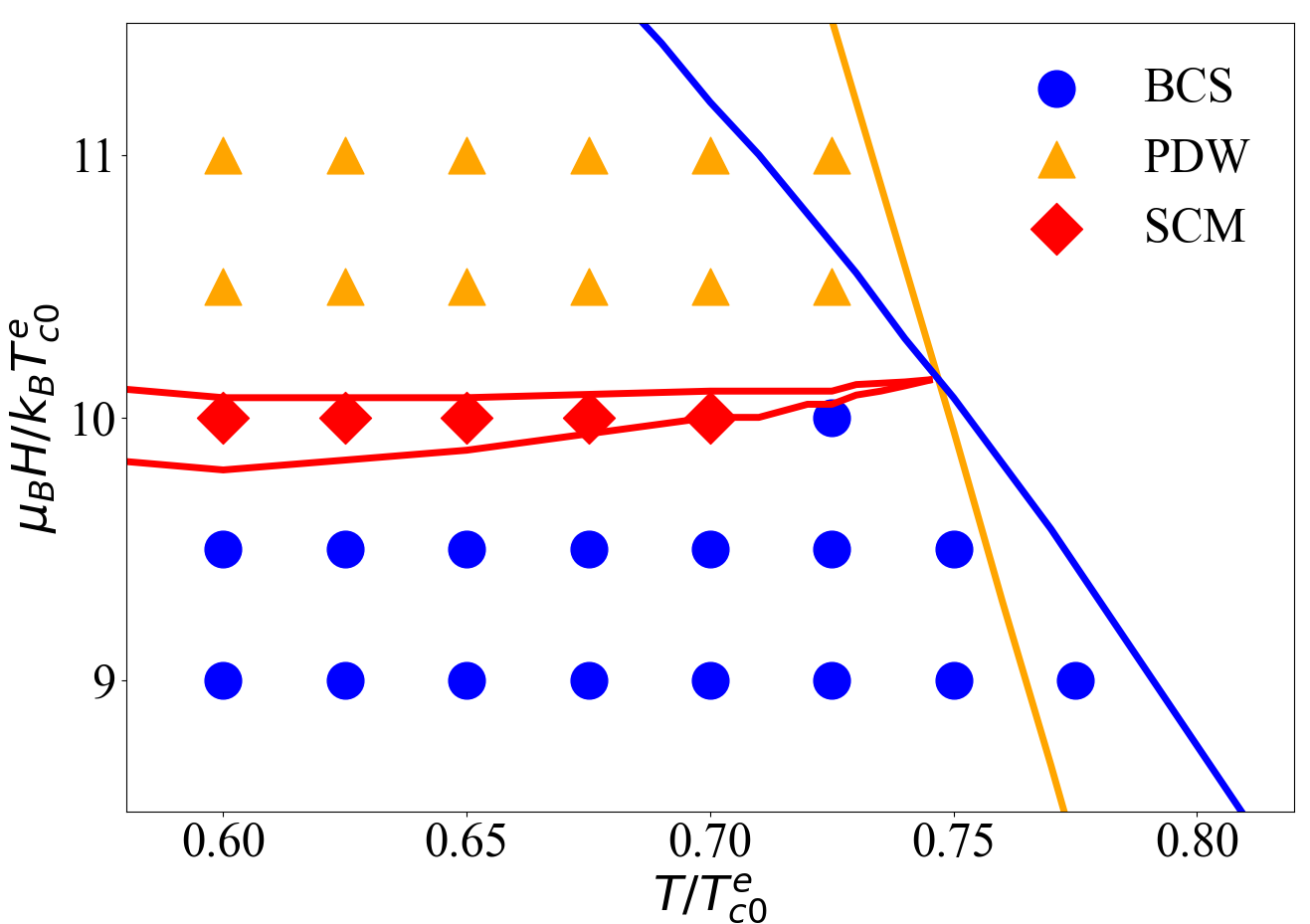}
  \caption{The phase diagram near the multicritical point for $\alpha/t_\perp=15.0$ and $\alpha_M=4.8$.
  Blue circles, orange triangles, and red diamonds represent the BCS, PDW, and SCM states, respectively.
  The SCM state is stable in the area between two phase transition (red) lines.
  The upper critical fields $H_{\rm c2}^{e,0}(T),\ H_{\rm c2}^{o,0}(T)$ are also depicted for reference.
  }
\label{fig:cond_energy_1}
\end{figure}

\begin{figure}[htbp]
  \centering
  \includegraphics[width=0.48\textwidth]{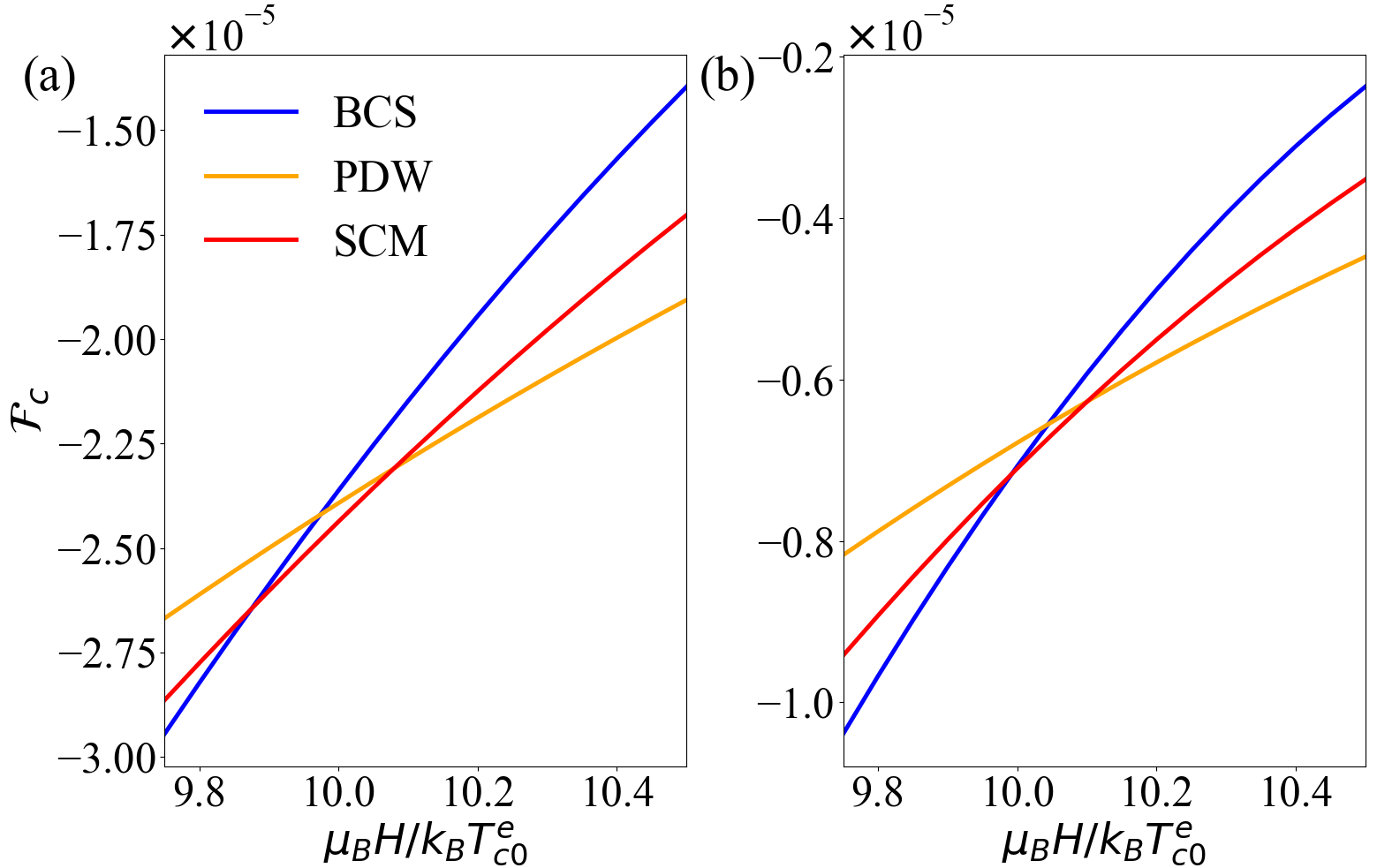}
  \caption{Magnetic field dependence of the condensation energy for the BCS (blue), PDW (orange), and SCM (red) states at (a) $T/T_{\mathrm{c}0}^{e}=0.65$ and (b) $T/T_{\mathrm{c}0}^{e}=0.70$.}
  \label{fig:h-dep_of_cond_energy}
\end{figure}

Near the multicritical point, the equilibrium vortex lattice states are determined by comparing the condensation energy Eq.~\eqref{eq:sk-1} for the BCS, PDW, and SCM states (Fig.~\ref{fig:cond_energy_1}).
The parity transition line can be drawn almost horizontally at around $\mu_{\rm B} H/k_{\rm B}T_{\mathrm{c}0}^{e}=10$ consistent with the experiment in $\mathrm{CeRh_2As_2}$~\cite{khim2021field}. 
Interestingly, the SCM state appears as an intermediate state between the low-field BCS and high-field PDW states.
Figure~\ref{fig:h-dep_of_cond_energy} shows the $H$-dependence of the condensation energy at fixed temperatures (a) $T/T_{\mathrm{c}0}^{e}=0.65$ and (b) $0.70$.
It can be seen that the SCM state has the minimum free energy in a small region between the BCS and PDW states.

The reason why the SCM state is stabilized can be understood by analyzing the GL free energy in real-space representation.
Near the multicritical point, the quadratic terms for both $\Delta_e$- and $\Delta_o$-channels vanish.
Then, the quartic terms determine the superconducting state.
In particular, the term proportional to $|\Delta_1(\bm{r})|^2|\Delta_2(\bm{r})|^2$ is essential, although this term was neglected in previous phenomenological studies~\cite{mockli2018orbitally,schertenleib2021unusual}. 
If the coefficient of this term is positive as expected, it can be interpreted as a repulsive force between vortices on adjacent layers.
Therefore, the SCM state where the vortex position is shifted between the neighboring layers is energetically favored.

\textit{Summary and discussion. ---}
Starting from the bilayer Rashba Hamiltonian as a minimal model of the LNC layered system with two sublattices, we derived the GL free energy functional microscopically.
By incorporating the vortex lattice formation, it was confirmed that the parity transition occurs between the BCS and PDW vortex phases, and the phase diagram of $\mathrm{CeRh_2As_2}$ is reproduced.
Furthermore, the presence of an intermediate phase called the superconducting meron phase is uncovered.
In this phase, the vortex structure breaks the parity symmetry and is characterized by the meron lattice pseudospin texture, an analog of the magnetic meron lattice. 
We emphasize that the SCM state is induced by the vortex degree of freedom, and has not been shown in theories of uniform states and even in the phenomenological models of vortex states.  
This result opens up the possibility that new superconducting states may arise from the combined sublattice and vortex degrees of freedom.

Experimentally, the SCM state can be verified by scanning tunneling microscopy, since the projection of vortex cores onto the $ab$ plane yields twice as many vortex cores as the BCS or PDW state. The vortex structure can also be investigated by small-angle neutron scattering and nuclear magnetic resonance measurements. 
Furthermore, a signature of the SCM state may appear in the transport and magnetization measurements since the vortex pinning is expected to be reduced as in the fractional vortex states of spin-triplet superconductors~\cite{Salomaa-Volovik}. The reduced pinning has recently been observed in a spin-triplet superconductor candidate UTe$_2$ by a critical current measurement~\cite{Tokiwa2023}. The magnetization curve of CeRh$_2$As$_2$~\cite{khim2021field} is consistent with the decrease of pinning around the multicritical point.

We are grateful to R. Ikeda, T. Hanaguri, K. Ishida, and S. Kitagawa for friutful discussions. This work was supported by JSPS KAKENHI (Grant Nos. JP21K18145, JP22H01181, JP22H04933, JP23K22452, JP23K17353, JP24H00007).

\bibliography{reference}

\begin{thebibliography}{73}%
\makeatletter
\providecommand \@ifxundefined [1]{%
 \@ifx{#1\undefined}
}%
\providecommand \@ifnum [1]{%
 \ifnum #1\expandafter \@firstoftwo
 \else \expandafter \@secondoftwo
 \fi
}%
\providecommand \@ifx [1]{%
 \ifx #1\expandafter \@firstoftwo
 \else \expandafter \@secondoftwo
 \fi
}%
\providecommand \natexlab [1]{#1}%
\providecommand \enquote  [1]{``#1''}%
\providecommand \bibnamefont  [1]{#1}%
\providecommand \bibfnamefont [1]{#1}%
\providecommand \citenamefont [1]{#1}%
\providecommand \href@noop [0]{\@secondoftwo}%
\providecommand \href [0]{\begingroup \@sanitize@url \@href}%
\providecommand \@href[1]{\@@startlink{#1}\@@href}%
\providecommand \@@href[1]{\endgroup#1\@@endlink}%
\providecommand \@sanitize@url [0]{\catcode `\\12\catcode `\$12\catcode `\&12\catcode `\#12\catcode `\^12\catcode `\_12\catcode `\%12\relax}%
\providecommand \@@startlink[1]{}%
\providecommand \@@endlink[0]{}%
\providecommand \url  [0]{\begingroup\@sanitize@url \@url }%
\providecommand \@url [1]{\endgroup\@href {#1}{\urlprefix }}%
\providecommand \urlprefix  [0]{URL }%
\providecommand \Eprint [0]{\href }%
\providecommand \doibase [0]{https://doi.org/}%
\providecommand \selectlanguage [0]{\@gobble}%
\providecommand \bibinfo  [0]{\@secondoftwo}%
\providecommand \bibfield  [0]{\@secondoftwo}%
\providecommand \translation [1]{[#1]}%
\providecommand \BibitemOpen [0]{}%
\providecommand \bibitemStop [0]{}%
\providecommand \bibitemNoStop [0]{.\EOS\space}%
\providecommand \EOS [0]{\spacefactor3000\relax}%
\providecommand \BibitemShut  [1]{\csname bibitem#1\endcsname}%
\let\auto@bib@innerbib\@empty
\bibitem [{\citenamefont {Steglich}\ \emph {et~al.}(1979)\citenamefont {Steglich}, \citenamefont {Aarts}, \citenamefont {Bredl}, \citenamefont {Lieke}, \citenamefont {Meschede}, \citenamefont {Franz},\ and\ \citenamefont {Sch{\"a}fer}}]{steglich1979superconductivity}%
  \BibitemOpen
  \bibfield  {author} {\bibinfo {author} {\bibfnamefont {F.}~\bibnamefont {Steglich}}, \bibinfo {author} {\bibfnamefont {J.}~\bibnamefont {Aarts}}, \bibinfo {author} {\bibfnamefont {C.}~\bibnamefont {Bredl}}, \bibinfo {author} {\bibfnamefont {W.}~\bibnamefont {Lieke}}, \bibinfo {author} {\bibfnamefont {D.}~\bibnamefont {Meschede}}, \bibinfo {author} {\bibfnamefont {W.}~\bibnamefont {Franz}},\ and\ \bibinfo {author} {\bibfnamefont {H.}~\bibnamefont {Sch{\"a}fer}},\ }\href {https://journals.aps.org/prl/abstract/10.1103/PhysRevLett.43.1892} {\bibfield  {journal} {\bibinfo  {journal} {Phys. Rev. Lett.}\ }\textbf {\bibinfo {volume} {43}},\ \bibinfo {pages} {1892} (\bibinfo {year} {1979})}\BibitemShut {NoStop}%
\bibitem [{\citenamefont {Bednorz}\ and\ \citenamefont {M{\"u}ller}(1986)}]{bednorz1986possible}%
  \BibitemOpen
  \bibfield  {author} {\bibinfo {author} {\bibfnamefont {J.~G.}\ \bibnamefont {Bednorz}}\ and\ \bibinfo {author} {\bibfnamefont {K.~A.}\ \bibnamefont {M{\"u}ller}},\ }\href {https://link.springer.com/article/10.1007/BF01303701} {\bibfield  {journal} {\bibinfo  {journal} {Z. Phys. B}\ }\textbf {\bibinfo {volume} {64}},\ \bibinfo {pages} {189} (\bibinfo {year} {1986})}\BibitemShut {NoStop}%
\bibitem [{\citenamefont {Bardeen}\ \emph {et~al.}(1957)\citenamefont {Bardeen}, \citenamefont {Cooper},\ and\ \citenamefont {Schrieffer}}]{bardeen1957microscopic}%
  \BibitemOpen
  \bibfield  {author} {\bibinfo {author} {\bibfnamefont {J.}~\bibnamefont {Bardeen}}, \bibinfo {author} {\bibfnamefont {L.~N.}\ \bibnamefont {Cooper}},\ and\ \bibinfo {author} {\bibfnamefont {J.~R.}\ \bibnamefont {Schrieffer}},\ }\href {https://doi.org/10.1103/PhysRev.106.162} {\bibfield  {journal} {\bibinfo  {journal} {Phys. Rev.}\ }\textbf {\bibinfo {volume} {106}},\ \bibinfo {pages} {162} (\bibinfo {year} {1957})}\BibitemShut {NoStop}%
\bibitem [{\citenamefont {M{\"o}ckli}(2022)}]{mockli2022unconventional}%
  \BibitemOpen
  \bibfield  {author} {\bibinfo {author} {\bibfnamefont {D.}~\bibnamefont {M{\"o}ckli}},\ }\href {https://iopscience.iop.org/article/10.1088/1742-6596/2164/1/012009} {\bibfield  {journal} {\bibinfo  {journal} {J. Phys.: Conf. Ser.}\ }\textbf {\bibinfo {volume} {2164}},\ \bibinfo {pages} {012009} (\bibinfo {year} {2022})}\BibitemShut {NoStop}%
\bibitem [{\citenamefont {Fischer}\ \emph {et~al.}(2023)\citenamefont {Fischer}, \citenamefont {Sigrist}, \citenamefont {Agterberg},\ and\ \citenamefont {Yanase}}]{fischer2023superconductivity}%
  \BibitemOpen
  \bibfield  {author} {\bibinfo {author} {\bibfnamefont {M.~H.}\ \bibnamefont {Fischer}}, \bibinfo {author} {\bibfnamefont {M.}~\bibnamefont {Sigrist}}, \bibinfo {author} {\bibfnamefont {D.~F.}\ \bibnamefont {Agterberg}},\ and\ \bibinfo {author} {\bibfnamefont {Y.}~\bibnamefont {Yanase}},\ }\href {https://www.annualreviews.org/content/journals/10.1146/annurev-conmatphys-040521-042511} {\bibfield  {journal} {\bibinfo  {journal} {Annu. Rev. Condens. Matter Phys.}\ }\textbf {\bibinfo {volume} {14}},\ \bibinfo {pages} {153} (\bibinfo {year} {2023})}\BibitemShut {NoStop}%
\bibitem [{\citenamefont {Aoki}\ and\ \citenamefont {Flouquet}(2011)}]{aoki2011ferromagnetism}%
  \BibitemOpen
  \bibfield  {author} {\bibinfo {author} {\bibfnamefont {D.}~\bibnamefont {Aoki}}\ and\ \bibinfo {author} {\bibfnamefont {J.}~\bibnamefont {Flouquet}},\ }\href {https://journals.jps.jp/doi/10.1143/JPSJ.81.011003} {\bibfield  {journal} {\bibinfo  {journal} {J. Phys. Soc. Jpn.}\ }\textbf {\bibinfo {volume} {81}},\ \bibinfo {pages} {011003} (\bibinfo {year} {2011})}\BibitemShut {NoStop}%
\bibitem [{\citenamefont {Aoki}\ \emph {et~al.}(2022)\citenamefont {Aoki}, \citenamefont {Brison}, \citenamefont {Flouquet}, \citenamefont {Ishida}, \citenamefont {Knebel}, \citenamefont {Tokunaga},\ and\ \citenamefont {Yanase}}]{aoki2022unconventional}%
  \BibitemOpen
  \bibfield  {author} {\bibinfo {author} {\bibfnamefont {D.}~\bibnamefont {Aoki}}, \bibinfo {author} {\bibfnamefont {J.~P.}\ \bibnamefont {Brison}}, \bibinfo {author} {\bibfnamefont {J.}~\bibnamefont {Flouquet}}, \bibinfo {author} {\bibfnamefont {K.}~\bibnamefont {Ishida}}, \bibinfo {author} {\bibfnamefont {G.}~\bibnamefont {Knebel}}, \bibinfo {author} {\bibfnamefont {Y.}~\bibnamefont {Tokunaga}},\ and\ \bibinfo {author} {\bibfnamefont {Y.}~\bibnamefont {Yanase}},\ }\href {https://iopscience.iop.org/article/10.1088/1361-648X/ac5863/meta} {\bibfield  {journal} {\bibinfo  {journal} {J. Phys. Condens. Matter}\ }\textbf {\bibinfo {volume} {34}},\ \bibinfo {pages} {243002} (\bibinfo {year} {2022})}\BibitemShut {NoStop}%
\bibitem [{\citenamefont {Nishikubo}\ \emph {et~al.}(2011)\citenamefont {Nishikubo}, \citenamefont {Kudo},\ and\ \citenamefont {Nohara}}]{nishikubo2011superconductivity}%
  \BibitemOpen
  \bibfield  {author} {\bibinfo {author} {\bibfnamefont {Y.}~\bibnamefont {Nishikubo}}, \bibinfo {author} {\bibfnamefont {K.}~\bibnamefont {Kudo}},\ and\ \bibinfo {author} {\bibfnamefont {M.}~\bibnamefont {Nohara}},\ }\href {https://journals.jps.jp/doi/10.1143/JPSJ.80.055002} {\bibfield  {journal} {\bibinfo  {journal} {J. Phys. Soc. Jpn.}\ }\textbf {\bibinfo {volume} {80}},\ \bibinfo {pages} {055002} (\bibinfo {year} {2011})}\BibitemShut {NoStop}%
\bibitem [{\citenamefont {Wilson}\ and\ \citenamefont {Yoffe}(1969)}]{wilson1969transition}%
  \BibitemOpen
  \bibfield  {author} {\bibinfo {author} {\bibfnamefont {J.~A.}\ \bibnamefont {Wilson}}\ and\ \bibinfo {author} {\bibfnamefont {A.}~\bibnamefont {Yoffe}},\ }\href {https://www.tandfonline.com/doi/abs/10.1080/00018736900101307} {\bibfield  {journal} {\bibinfo  {journal} {Adv. Phys.}\ }\textbf {\bibinfo {volume} {18}},\ \bibinfo {pages} {193} (\bibinfo {year} {1969})}\BibitemShut {NoStop}%
\bibitem [{\citenamefont {Mukuda}\ \emph {et~al.}(2011)\citenamefont {Mukuda}, \citenamefont {Shimizu}, \citenamefont {Iyo},\ and\ \citenamefont {Kitaoka}}]{mukuda2011high}%
  \BibitemOpen
  \bibfield  {author} {\bibinfo {author} {\bibfnamefont {H.}~\bibnamefont {Mukuda}}, \bibinfo {author} {\bibfnamefont {S.}~\bibnamefont {Shimizu}}, \bibinfo {author} {\bibfnamefont {A.}~\bibnamefont {Iyo}},\ and\ \bibinfo {author} {\bibfnamefont {Y.}~\bibnamefont {Kitaoka}},\ }\href {https://journals.jps.jp/doi/10.1143/JPSJ.81.011008} {\bibfield  {journal} {\bibinfo  {journal} {J. Phys. Soc. Jpn.}\ }\textbf {\bibinfo {volume} {81}},\ \bibinfo {pages} {011008} (\bibinfo {year} {2011})}\BibitemShut {NoStop}%
\bibitem [{\citenamefont {Mizukami}\ \emph {et~al.}(2011)\citenamefont {Mizukami}, \citenamefont {Shishido}, \citenamefont {Shibauchi}, \citenamefont {Shimozawa}, \citenamefont {Yasumoto}, \citenamefont {Watanabe}, \citenamefont {Yamashita}, \citenamefont {Ikeda}, \citenamefont {Terashima}, \citenamefont {Kontani},\ and\ \citenamefont {Matsuda}}]{mizukami2011extremely}%
  \BibitemOpen
  \bibfield  {author} {\bibinfo {author} {\bibfnamefont {Y.}~\bibnamefont {Mizukami}}, \bibinfo {author} {\bibfnamefont {H.}~\bibnamefont {Shishido}}, \bibinfo {author} {\bibfnamefont {T.}~\bibnamefont {Shibauchi}}, \bibinfo {author} {\bibfnamefont {M.}~\bibnamefont {Shimozawa}}, \bibinfo {author} {\bibfnamefont {S.}~\bibnamefont {Yasumoto}}, \bibinfo {author} {\bibfnamefont {D.}~\bibnamefont {Watanabe}}, \bibinfo {author} {\bibfnamefont {M.}~\bibnamefont {Yamashita}}, \bibinfo {author} {\bibfnamefont {H.}~\bibnamefont {Ikeda}}, \bibinfo {author} {\bibfnamefont {T.}~\bibnamefont {Terashima}}, \bibinfo {author} {\bibfnamefont {H.}~\bibnamefont {Kontani}},\ and\ \bibinfo {author} {\bibfnamefont {Y.}~\bibnamefont {Matsuda}},\ }\href {https://www.nature.com/articles/nphys2112} {\bibfield  {journal} {\bibinfo  {journal} {Nature physics}\ }\textbf {\bibinfo {volume} {7}},\ \bibinfo {pages} {849} (\bibinfo {year} {2011})}\BibitemShut {NoStop}%
\bibitem [{\citenamefont {Yoshida}\ \emph {et~al.}(2012)\citenamefont {Yoshida}, \citenamefont {Sigrist},\ and\ \citenamefont {Yanase}}]{yoshida2012pair}%
  \BibitemOpen
  \bibfield  {author} {\bibinfo {author} {\bibfnamefont {T.}~\bibnamefont {Yoshida}}, \bibinfo {author} {\bibfnamefont {M.}~\bibnamefont {Sigrist}},\ and\ \bibinfo {author} {\bibfnamefont {Y.}~\bibnamefont {Yanase}},\ }\href {https://journals.aps.org/prb/abstract/10.1103/PhysRevB.86.134514} {\bibfield  {journal} {\bibinfo  {journal} {Phys. Rev. B}\ }\textbf {\bibinfo {volume} {86}},\ \bibinfo {pages} {134514} (\bibinfo {year} {2012})}\BibitemShut {NoStop}%
\bibitem [{\citenamefont {Yoshida}\ \emph {et~al.}(2015)\citenamefont {Yoshida}, \citenamefont {Sigrist},\ and\ \citenamefont {Yanase}}]{yoshida2015topological}%
  \BibitemOpen
  \bibfield  {author} {\bibinfo {author} {\bibfnamefont {T.}~\bibnamefont {Yoshida}}, \bibinfo {author} {\bibfnamefont {M.}~\bibnamefont {Sigrist}},\ and\ \bibinfo {author} {\bibfnamefont {Y.}~\bibnamefont {Yanase}},\ }\href {https://journals.aps.org/prl/abstract/10.1103/PhysRevLett.115.027001} {\bibfield  {journal} {\bibinfo  {journal} {Phys. Rev. Lett.}\ }\textbf {\bibinfo {volume} {115}},\ \bibinfo {pages} {027001} (\bibinfo {year} {2015})}\BibitemShut {NoStop}%
\bibitem [{\citenamefont {Nogaki}\ \emph {et~al.}(2021)\citenamefont {Nogaki}, \citenamefont {Daido}, \citenamefont {Ishizuka},\ and\ \citenamefont {Yanase}}]{nogaki2021topological}%
  \BibitemOpen
  \bibfield  {author} {\bibinfo {author} {\bibfnamefont {K.}~\bibnamefont {Nogaki}}, \bibinfo {author} {\bibfnamefont {A.}~\bibnamefont {Daido}}, \bibinfo {author} {\bibfnamefont {J.}~\bibnamefont {Ishizuka}},\ and\ \bibinfo {author} {\bibfnamefont {Y.}~\bibnamefont {Yanase}},\ }\href {https://journals.aps.org/prresearch/abstract/10.1103/PhysRevResearch.3.L032071} {\bibfield  {journal} {\bibinfo  {journal} {Phys. Rev. Research}\ }\textbf {\bibinfo {volume} {3}},\ \bibinfo {pages} {L032071} (\bibinfo {year} {2021})}\BibitemShut {NoStop}%
\bibitem [{\citenamefont {Ishizuka}\ \emph {et~al.}(2023)\citenamefont {Ishizuka}, \citenamefont {Nogaki}, \citenamefont {Sigrist},\ and\ \citenamefont {Yanase}}]{ishizuka2023correlation}%
  \BibitemOpen
  \bibfield  {author} {\bibinfo {author} {\bibfnamefont {J.}~\bibnamefont {Ishizuka}}, \bibinfo {author} {\bibfnamefont {K.}~\bibnamefont {Nogaki}}, \bibinfo {author} {\bibfnamefont {M.}~\bibnamefont {Sigrist}},\ and\ \bibinfo {author} {\bibfnamefont {Y.}~\bibnamefont {Yanase}},\ }\href {https://arxiv.org/abs/2311.00324} {\bibfield  {journal} {\bibinfo  {journal} {arXiv:2311.00324 [cond-mat.supr-con]}\ } (\bibinfo {year} {2023})}\BibitemShut {NoStop}%
\bibitem [{\citenamefont {Yoshida}\ \emph {et~al.}(2014)\citenamefont {Yoshida}, \citenamefont {Sigrist},\ and\ \citenamefont {Yanase}}]{yoshida2014parity}%
  \BibitemOpen
  \bibfield  {author} {\bibinfo {author} {\bibfnamefont {T.}~\bibnamefont {Yoshida}}, \bibinfo {author} {\bibfnamefont {M.}~\bibnamefont {Sigrist}},\ and\ \bibinfo {author} {\bibfnamefont {Y.}~\bibnamefont {Yanase}},\ }\href {https://journals.jps.jp/doi/10.7566/JPSJ.83.013703} {\bibfield  {journal} {\bibinfo  {journal} {J. Phys. Soc. Jpn.}\ }\textbf {\bibinfo {volume} {83}},\ \bibinfo {pages} {013703} (\bibinfo {year} {2014})}\BibitemShut {NoStop}%
\bibitem [{\citenamefont {Sigrist}\ \emph {et~al.}(2014)\citenamefont {Sigrist}, \citenamefont {Agterberg}, \citenamefont {Fischer}, \citenamefont {Goryo}, \citenamefont {Loder}, \citenamefont {Rhim}, \citenamefont {Maruyama}, \citenamefont {Yanase}, \citenamefont {Yoshida},\ and\ \citenamefont {Youn}}]{sigrist2014superconductors}%
  \BibitemOpen
  \bibfield  {author} {\bibinfo {author} {\bibfnamefont {M.}~\bibnamefont {Sigrist}}, \bibinfo {author} {\bibfnamefont {D.~F.}\ \bibnamefont {Agterberg}}, \bibinfo {author} {\bibfnamefont {M.~H.}\ \bibnamefont {Fischer}}, \bibinfo {author} {\bibfnamefont {J.}~\bibnamefont {Goryo}}, \bibinfo {author} {\bibfnamefont {F.}~\bibnamefont {Loder}}, \bibinfo {author} {\bibfnamefont {S.-H.}\ \bibnamefont {Rhim}}, \bibinfo {author} {\bibfnamefont {D.}~\bibnamefont {Maruyama}}, \bibinfo {author} {\bibfnamefont {Y.}~\bibnamefont {Yanase}}, \bibinfo {author} {\bibfnamefont {T.}~\bibnamefont {Yoshida}},\ and\ \bibinfo {author} {\bibfnamefont {S.~J.}\ \bibnamefont {Youn}},\ }\href {https://journals.jps.jp/doi/10.7566/JPSJ.83.061014} {\bibfield  {journal} {\bibinfo  {journal} {J. Phys. Soc. Jpn.}\ }\textbf {\bibinfo {volume} {83}},\ \bibinfo {pages} {061014} (\bibinfo {year} {2014})}\BibitemShut {NoStop}%
\bibitem [{\citenamefont {Higashi}\ \emph {et~al.}(2016)\citenamefont {Higashi}, \citenamefont {Nagai}, \citenamefont {Yoshida}, \citenamefont {Masaki},\ and\ \citenamefont {Yanase}}]{higashi2016robust}%
  \BibitemOpen
  \bibfield  {author} {\bibinfo {author} {\bibfnamefont {Y.}~\bibnamefont {Higashi}}, \bibinfo {author} {\bibfnamefont {Y.}~\bibnamefont {Nagai}}, \bibinfo {author} {\bibfnamefont {T.}~\bibnamefont {Yoshida}}, \bibinfo {author} {\bibfnamefont {Y.}~\bibnamefont {Masaki}},\ and\ \bibinfo {author} {\bibfnamefont {Y.}~\bibnamefont {Yanase}},\ }\href {https://journals.aps.org/prb/abstract/10.1103/PhysRevB.93.104529} {\bibfield  {journal} {\bibinfo  {journal} {Phys. Rev. B}\ }\textbf {\bibinfo {volume} {93}},\ \bibinfo {pages} {104529} (\bibinfo {year} {2016})}\BibitemShut {NoStop}%
\bibitem [{\citenamefont {Skurativska}\ \emph {et~al.}(2021)\citenamefont {Skurativska}, \citenamefont {Sigrist},\ and\ \citenamefont {Fischer}}]{skurativska2021spin}%
  \BibitemOpen
  \bibfield  {author} {\bibinfo {author} {\bibfnamefont {A.}~\bibnamefont {Skurativska}}, \bibinfo {author} {\bibfnamefont {M.}~\bibnamefont {Sigrist}},\ and\ \bibinfo {author} {\bibfnamefont {M.~H.}\ \bibnamefont {Fischer}},\ }\href {https://journals.aps.org/prresearch/abstract/10.1103/PhysRevResearch.3.033133} {\bibfield  {journal} {\bibinfo  {journal} {Phys. Rev. Research}\ }\textbf {\bibinfo {volume} {3}},\ \bibinfo {pages} {033133} (\bibinfo {year} {2021})}\BibitemShut {NoStop}%
\bibitem [{\citenamefont {Nogaki}\ and\ \citenamefont {Yanase}(2022)}]{nogaki2022even}%
  \BibitemOpen
  \bibfield  {author} {\bibinfo {author} {\bibfnamefont {K.}~\bibnamefont {Nogaki}}\ and\ \bibinfo {author} {\bibfnamefont {Y.}~\bibnamefont {Yanase}},\ }\href {https://journals.aps.org/prb/abstract/10.1103/PhysRevB.106.L100504} {\bibfield  {journal} {\bibinfo  {journal} {Phys. Rev. B}\ }\textbf {\bibinfo {volume} {106}},\ \bibinfo {pages} {L100504} (\bibinfo {year} {2022})}\BibitemShut {NoStop}%
\bibitem [{\citenamefont {Hackner}\ and\ \citenamefont {Brydon}(2023)}]{hackner2023bardasis}%
  \BibitemOpen
  \bibfield  {author} {\bibinfo {author} {\bibfnamefont {N.~A.}\ \bibnamefont {Hackner}}\ and\ \bibinfo {author} {\bibfnamefont {P.}~\bibnamefont {Brydon}},\ }\href {https://journals.aps.org/prb/abstract/10.1103/PhysRevB.108.L220505} {\bibfield  {journal} {\bibinfo  {journal} {Phys. Rev. B}\ }\textbf {\bibinfo {volume} {108}},\ \bibinfo {pages} {L220505} (\bibinfo {year} {2023})}\BibitemShut {NoStop}%
\bibitem [{\citenamefont {Lee}\ and\ \citenamefont {Chung}(2023)}]{lee2023linear}%
  \BibitemOpen
  \bibfield  {author} {\bibinfo {author} {\bibfnamefont {C.}~\bibnamefont {Lee}}\ and\ \bibinfo {author} {\bibfnamefont {S.~B.}\ \bibnamefont {Chung}},\ }\href {https://www.nature.com/articles/s42005-023-01421-8} {\bibfield  {journal} {\bibinfo  {journal} {Commun. Phys.}\ }\textbf {\bibinfo {volume} {6}},\ \bibinfo {pages} {307} (\bibinfo {year} {2023})}\BibitemShut {NoStop}%
\bibitem [{\citenamefont {Szab{\'o}}\ and\ \citenamefont {Ramires}(2023)}]{szabo2023superconductivity}%
  \BibitemOpen
  \bibfield  {author} {\bibinfo {author} {\bibfnamefont {A.~L.}\ \bibnamefont {Szab{\'o}}}\ and\ \bibinfo {author} {\bibfnamefont {A.}~\bibnamefont {Ramires}},\ }\href {https://arxiv.org/abs/2309.05664} {\bibfield  {journal} {\bibinfo  {journal} {arXiv:2309.05664 [cond-mat.supr-con]}\ } (\bibinfo {year} {2023})}\BibitemShut {NoStop}%
\bibitem [{\citenamefont {Nogaki}\ and\ \citenamefont {Yanase}(2023)}]{nogaki2023field}%
  \BibitemOpen
  \bibfield  {author} {\bibinfo {author} {\bibfnamefont {K.}~\bibnamefont {Nogaki}}\ and\ \bibinfo {author} {\bibfnamefont {Y.}~\bibnamefont {Yanase}},\ }\href {https://arxiv.org/abs/2312.07053} {\bibfield  {journal} {\bibinfo  {journal} {arXiv:2312.07053 [cond-mat.supr-con]}\ } (\bibinfo {year} {2023})}\BibitemShut {NoStop}%
\bibitem [{\citenamefont {Szab{\'o}}\ \emph {et~al.}(2024)\citenamefont {Szab{\'o}}, \citenamefont {Fischer},\ and\ \citenamefont {Sigrist}}]{szabo2024effects}%
  \BibitemOpen
  \bibfield  {author} {\bibinfo {author} {\bibfnamefont {A.~L.}\ \bibnamefont {Szab{\'o}}}, \bibinfo {author} {\bibfnamefont {M.~H.}\ \bibnamefont {Fischer}},\ and\ \bibinfo {author} {\bibfnamefont {M.}~\bibnamefont {Sigrist}},\ }\href {https://journals.aps.org/prresearch/abstract/10.1103/PhysRevResearch.6.023080} {\bibfield  {journal} {\bibinfo  {journal} {Physical Review Research}\ }\textbf {\bibinfo {volume} {6}},\ \bibinfo {pages} {023080} (\bibinfo {year} {2024})}\BibitemShut {NoStop}%
\bibitem [{\citenamefont {Fischer}\ \emph {et~al.}(2011)\citenamefont {Fischer}, \citenamefont {Loder},\ and\ \citenamefont {Sigrist}}]{fischer2011superconductivity}%
  \BibitemOpen
  \bibfield  {author} {\bibinfo {author} {\bibfnamefont {M.~H.}\ \bibnamefont {Fischer}}, \bibinfo {author} {\bibfnamefont {F.}~\bibnamefont {Loder}},\ and\ \bibinfo {author} {\bibfnamefont {M.}~\bibnamefont {Sigrist}},\ }\href {https://journals.aps.org/prb/abstract/10.1103/PhysRevB.84.184533} {\bibfield  {journal} {\bibinfo  {journal} {Phys. Rev. B}\ }\textbf {\bibinfo {volume} {84}},\ \bibinfo {pages} {184533} (\bibinfo {year} {2011})}\BibitemShut {NoStop}%
\bibitem [{\citenamefont {Watanabe}\ \emph {et~al.}(2015)\citenamefont {Watanabe}, \citenamefont {Yoshida},\ and\ \citenamefont {Yanase}}]{watanabe2015odd}%
  \BibitemOpen
  \bibfield  {author} {\bibinfo {author} {\bibfnamefont {T.}~\bibnamefont {Watanabe}}, \bibinfo {author} {\bibfnamefont {T.}~\bibnamefont {Yoshida}},\ and\ \bibinfo {author} {\bibfnamefont {Y.}~\bibnamefont {Yanase}},\ }\href {https://journals.aps.org/prb/abstract/10.1103/PhysRevB.92.174502} {\bibfield  {journal} {\bibinfo  {journal} {Phys. Rev. B}\ }\textbf {\bibinfo {volume} {92}},\ \bibinfo {pages} {174502} (\bibinfo {year} {2015})}\BibitemShut {NoStop}%
\bibitem [{\citenamefont {Sumita}\ and\ \citenamefont {Yanase}(2016)}]{sumita2016superconductivity}%
  \BibitemOpen
  \bibfield  {author} {\bibinfo {author} {\bibfnamefont {S.}~\bibnamefont {Sumita}}\ and\ \bibinfo {author} {\bibfnamefont {Y.}~\bibnamefont {Yanase}},\ }\href {https://journals.aps.org/prb/abstract/10.1103/PhysRevB.93.224507} {\bibfield  {journal} {\bibinfo  {journal} {Phys. Rev. B}\ }\textbf {\bibinfo {volume} {93}},\ \bibinfo {pages} {224507} (\bibinfo {year} {2016})}\BibitemShut {NoStop}%
\bibitem [{\citenamefont {Nakamura}\ and\ \citenamefont {Yanase}(2017)}]{nakamura2017odd}%
  \BibitemOpen
  \bibfield  {author} {\bibinfo {author} {\bibfnamefont {Y.}~\bibnamefont {Nakamura}}\ and\ \bibinfo {author} {\bibfnamefont {Y.}~\bibnamefont {Yanase}},\ }\href {https://journals.aps.org/prb/abstract/10.1103/PhysRevB.96.054501} {\bibfield  {journal} {\bibinfo  {journal} {Phys. Rev. B}\ }\textbf {\bibinfo {volume} {96}},\ \bibinfo {pages} {054501} (\bibinfo {year} {2017})}\BibitemShut {NoStop}%
\bibitem [{\citenamefont {Khim}\ \emph {et~al.}(2021)\citenamefont {Khim}, \citenamefont {Landaeta}, \citenamefont {Banda}, \citenamefont {Bannor}, \citenamefont {Brando}, \citenamefont {Brydon}, \citenamefont {Hafner}, \citenamefont {K{\"u}chler}, \citenamefont {Cardoso-Gil}, \citenamefont {Stockert}, \citenamefont {Mackenzie}, \citenamefont {Agterberg}, \citenamefont {Geibel},\ and\ \citenamefont {Hassinger}}]{khim2021field}%
  \BibitemOpen
  \bibfield  {author} {\bibinfo {author} {\bibfnamefont {S.}~\bibnamefont {Khim}}, \bibinfo {author} {\bibfnamefont {J.}~\bibnamefont {Landaeta}}, \bibinfo {author} {\bibfnamefont {J.}~\bibnamefont {Banda}}, \bibinfo {author} {\bibfnamefont {N.}~\bibnamefont {Bannor}}, \bibinfo {author} {\bibfnamefont {M.}~\bibnamefont {Brando}}, \bibinfo {author} {\bibfnamefont {P.}~\bibnamefont {Brydon}}, \bibinfo {author} {\bibfnamefont {D.}~\bibnamefont {Hafner}}, \bibinfo {author} {\bibfnamefont {R.}~\bibnamefont {K{\"u}chler}}, \bibinfo {author} {\bibfnamefont {R.}~\bibnamefont {Cardoso-Gil}}, \bibinfo {author} {\bibfnamefont {U.}~\bibnamefont {Stockert}}, \bibinfo {author} {\bibfnamefont {A.~P.}\ \bibnamefont {Mackenzie}}, \bibinfo {author} {\bibfnamefont {D.~F.}\ \bibnamefont {Agterberg}}, \bibinfo {author} {\bibfnamefont {C.}~\bibnamefont {Geibel}},\ and\ \bibinfo {author} {\bibfnamefont {E.}~\bibnamefont {Hassinger}},\ }\href {https://www.science.org/doi/10.1126/science.abe7518} {\bibfield  {journal} {\bibinfo
  {journal} {Science}\ }\textbf {\bibinfo {volume} {373}},\ \bibinfo {pages} {1012} (\bibinfo {year} {2021})}\BibitemShut {NoStop}%
\bibitem [{\citenamefont {Kimura}\ \emph {et~al.}(2021)\citenamefont {Kimura}, \citenamefont {Sichelschmidt},\ and\ \citenamefont {Khim}}]{kimura2021optical}%
  \BibitemOpen
  \bibfield  {author} {\bibinfo {author} {\bibfnamefont {S.}~\bibnamefont {Kimura}}, \bibinfo {author} {\bibfnamefont {J.}~\bibnamefont {Sichelschmidt}},\ and\ \bibinfo {author} {\bibfnamefont {S.}~\bibnamefont {Khim}},\ }\href {https://journals.aps.org/prb/abstract/10.1103/PhysRevB.104.245116} {\bibfield  {journal} {\bibinfo  {journal} {Phys. Rev. B}\ }\textbf {\bibinfo {volume} {104}},\ \bibinfo {pages} {245116} (\bibinfo {year} {2021})}\BibitemShut {NoStop}%
\bibitem [{\citenamefont {Onishi}\ \emph {et~al.}(2022)\citenamefont {Onishi}, \citenamefont {Stockert}, \citenamefont {Khim}, \citenamefont {Banda}, \citenamefont {Brando},\ and\ \citenamefont {Hassinger}}]{onishi2022low}%
  \BibitemOpen
  \bibfield  {author} {\bibinfo {author} {\bibfnamefont {S.}~\bibnamefont {Onishi}}, \bibinfo {author} {\bibfnamefont {U.}~\bibnamefont {Stockert}}, \bibinfo {author} {\bibfnamefont {S.}~\bibnamefont {Khim}}, \bibinfo {author} {\bibfnamefont {J.}~\bibnamefont {Banda}}, \bibinfo {author} {\bibfnamefont {M.}~\bibnamefont {Brando}},\ and\ \bibinfo {author} {\bibfnamefont {E.}~\bibnamefont {Hassinger}},\ }\href {https://www.frontiersin.org/articles/10.3389/femat.2022.880579/full} {\bibfield  {journal} {\bibinfo  {journal} {Frontiers in Electronic Materials}\ }\textbf {\bibinfo {volume} {2}},\ \bibinfo {pages} {880579} (\bibinfo {year} {2022})}\BibitemShut {NoStop}%
\bibitem [{\citenamefont {Landaeta}\ \emph {et~al.}(2022)\citenamefont {Landaeta}, \citenamefont {Khanenko}, \citenamefont {Cavanagh}, \citenamefont {Geibel}, \citenamefont {Khim}, \citenamefont {Mishra}, \citenamefont {Sheikin}, \citenamefont {Brydon}, \citenamefont {Agterberg}, \citenamefont {Brando},\ and\ \citenamefont {Hassinger}}]{landaeta2022field}%
  \BibitemOpen
  \bibfield  {author} {\bibinfo {author} {\bibfnamefont {J.}~\bibnamefont {Landaeta}}, \bibinfo {author} {\bibfnamefont {P.}~\bibnamefont {Khanenko}}, \bibinfo {author} {\bibfnamefont {D.}~\bibnamefont {Cavanagh}}, \bibinfo {author} {\bibfnamefont {C.}~\bibnamefont {Geibel}}, \bibinfo {author} {\bibfnamefont {S.}~\bibnamefont {Khim}}, \bibinfo {author} {\bibfnamefont {S.}~\bibnamefont {Mishra}}, \bibinfo {author} {\bibfnamefont {I.}~\bibnamefont {Sheikin}}, \bibinfo {author} {\bibfnamefont {P.}~\bibnamefont {Brydon}}, \bibinfo {author} {\bibfnamefont {D.}~\bibnamefont {Agterberg}}, \bibinfo {author} {\bibfnamefont {M.}~\bibnamefont {Brando}},\ and\ \bibinfo {author} {\bibfnamefont {E.}~\bibnamefont {Hassinger}},\ }\href {https://journals.aps.org/prx/abstract/10.1103/PhysRevX.12.031001} {\bibfield  {journal} {\bibinfo  {journal} {Phys. Rev. X}\ }\textbf {\bibinfo {volume} {12}},\ \bibinfo {pages} {031001} (\bibinfo {year} {2022})}\BibitemShut {NoStop}%
\bibitem [{\citenamefont {Mishra}\ \emph {et~al.}(2022)\citenamefont {Mishra}, \citenamefont {Liu}, \citenamefont {Bauer}, \citenamefont {Ronning},\ and\ \citenamefont {Thomas}}]{mishra2022anisotropic}%
  \BibitemOpen
  \bibfield  {author} {\bibinfo {author} {\bibfnamefont {S.}~\bibnamefont {Mishra}}, \bibinfo {author} {\bibfnamefont {Y.}~\bibnamefont {Liu}}, \bibinfo {author} {\bibfnamefont {E.~D.}\ \bibnamefont {Bauer}}, \bibinfo {author} {\bibfnamefont {F.}~\bibnamefont {Ronning}},\ and\ \bibinfo {author} {\bibfnamefont {S.~M.}\ \bibnamefont {Thomas}},\ }\href {https://journals.aps.org/prb/abstract/10.1103/PhysRevB.106.L140502} {\bibfield  {journal} {\bibinfo  {journal} {Phys. Rev. B}\ }\textbf {\bibinfo {volume} {106}},\ \bibinfo {pages} {L140502} (\bibinfo {year} {2022})}\BibitemShut {NoStop}%
\bibitem [{\citenamefont {Semeniuk}\ \emph {et~al.}(2023)\citenamefont {Semeniuk}, \citenamefont {Hafner}, \citenamefont {Khanenko}, \citenamefont {L{\"u}hmann}, \citenamefont {Banda}, \citenamefont {Landaeta}, \citenamefont {Geibel}, \citenamefont {Khim}, \citenamefont {Hassinger},\ and\ \citenamefont {Brando}}]{semeniuk2023decoupling}%
  \BibitemOpen
  \bibfield  {author} {\bibinfo {author} {\bibfnamefont {K.}~\bibnamefont {Semeniuk}}, \bibinfo {author} {\bibfnamefont {D.}~\bibnamefont {Hafner}}, \bibinfo {author} {\bibfnamefont {P.}~\bibnamefont {Khanenko}}, \bibinfo {author} {\bibfnamefont {T.}~\bibnamefont {L{\"u}hmann}}, \bibinfo {author} {\bibfnamefont {J.}~\bibnamefont {Banda}}, \bibinfo {author} {\bibfnamefont {J.~F.}\ \bibnamefont {Landaeta}}, \bibinfo {author} {\bibfnamefont {C.}~\bibnamefont {Geibel}}, \bibinfo {author} {\bibfnamefont {S.}~\bibnamefont {Khim}}, \bibinfo {author} {\bibfnamefont {E.}~\bibnamefont {Hassinger}},\ and\ \bibinfo {author} {\bibfnamefont {M.}~\bibnamefont {Brando}},\ }\href {https://journals.aps.org/prb/abstract/10.1103/PhysRevB.107.L220504} {\bibfield  {journal} {\bibinfo  {journal} {Phys. Rev. B}\ }\textbf {\bibinfo {volume} {107}},\ \bibinfo {pages} {L220504} (\bibinfo {year} {2023})}\BibitemShut {NoStop}%
\bibitem [{\citenamefont {Siddiquee}\ \emph {et~al.}(2023)\citenamefont {Siddiquee}, \citenamefont {Rehfuss}, \citenamefont {Broyles},\ and\ \citenamefont {Ran}}]{siddiquee2023pressure}%
  \BibitemOpen
  \bibfield  {author} {\bibinfo {author} {\bibfnamefont {H.}~\bibnamefont {Siddiquee}}, \bibinfo {author} {\bibfnamefont {Z.}~\bibnamefont {Rehfuss}}, \bibinfo {author} {\bibfnamefont {C.}~\bibnamefont {Broyles}},\ and\ \bibinfo {author} {\bibfnamefont {S.}~\bibnamefont {Ran}},\ }\href {https://journals.aps.org/prb/abstract/10.1103/PhysRevB.108.L020504} {\bibfield  {journal} {\bibinfo  {journal} {Phys. Rev. B}\ }\textbf {\bibinfo {volume} {108}},\ \bibinfo {pages} {L020504} (\bibinfo {year} {2023})}\BibitemShut {NoStop}%
\bibitem [{\citenamefont {Chen}\ \emph {et~al.}(2023)\citenamefont {Chen}, \citenamefont {Wang}, \citenamefont {Ishizuka}, \citenamefont {Nogaki}, \citenamefont {Cheng}, \citenamefont {Yang}, \citenamefont {Zhang}, \citenamefont {Chen}, \citenamefont {Zhu}, \citenamefont {Yanase}, \citenamefont {Lv},\ and\ \citenamefont {Huang}}]{chen2023coexistence}%
  \BibitemOpen
  \bibfield  {author} {\bibinfo {author} {\bibfnamefont {X.}~\bibnamefont {Chen}}, \bibinfo {author} {\bibfnamefont {L.}~\bibnamefont {Wang}}, \bibinfo {author} {\bibfnamefont {J.}~\bibnamefont {Ishizuka}}, \bibinfo {author} {\bibfnamefont {K.}~\bibnamefont {Nogaki}}, \bibinfo {author} {\bibfnamefont {Y.}~\bibnamefont {Cheng}}, \bibinfo {author} {\bibfnamefont {F.}~\bibnamefont {Yang}}, \bibinfo {author} {\bibfnamefont {R.}~\bibnamefont {Zhang}}, \bibinfo {author} {\bibfnamefont {Z.}~\bibnamefont {Chen}}, \bibinfo {author} {\bibfnamefont {F.}~\bibnamefont {Zhu}}, \bibinfo {author} {\bibfnamefont {Y.}~\bibnamefont {Yanase}}, \bibinfo {author} {\bibfnamefont {B.}~\bibnamefont {Lv}},\ and\ \bibinfo {author} {\bibfnamefont {Y.}~\bibnamefont {Huang}},\ }\href {https://arxiv.org/abs/2309.05895} {\bibfield  {journal} {\bibinfo  {journal} {arXiv:2309.05895 [cond-mat.str-el]}\ } (\bibinfo {year} {2023})}\BibitemShut {NoStop}%
\bibitem [{\citenamefont {Wu}\ \emph {et~al.}(2023)\citenamefont {Wu}, \citenamefont {Zhang}, \citenamefont {Ju}, \citenamefont {Hu}, \citenamefont {Huang}, \citenamefont {Zhang}, \citenamefont {Zhang}, \citenamefont {Zheng}, \citenamefont {Yang}, \citenamefont {Eljaouhari}, \citenamefont {Song}, \citenamefont {C.~Plumb}, \citenamefont {Steglich}, \citenamefont {Shi}, \citenamefont {Zwicknagi}, \citenamefont {Cao}, \citenamefont {Yuan},\ and\ \citenamefont {Liu}}]{wu2023fermi}%
  \BibitemOpen
  \bibfield  {author} {\bibinfo {author} {\bibfnamefont {Y.}~\bibnamefont {Wu}}, \bibinfo {author} {\bibfnamefont {Y.}~\bibnamefont {Zhang}}, \bibinfo {author} {\bibfnamefont {S.}~\bibnamefont {Ju}}, \bibinfo {author} {\bibfnamefont {Y.}~\bibnamefont {Hu}}, \bibinfo {author} {\bibfnamefont {Y.}~\bibnamefont {Huang}}, \bibinfo {author} {\bibfnamefont {Y.}~\bibnamefont {Zhang}}, \bibinfo {author} {\bibfnamefont {H.}~\bibnamefont {Zhang}}, \bibinfo {author} {\bibfnamefont {H.}~\bibnamefont {Zheng}}, \bibinfo {author} {\bibfnamefont {G.}~\bibnamefont {Yang}}, \bibinfo {author} {\bibfnamefont {E.-O.}\ \bibnamefont {Eljaouhari}}, \bibinfo {author} {\bibfnamefont {B.}~\bibnamefont {Song}}, \bibinfo {author} {\bibfnamefont {N.}~\bibnamefont {C.~Plumb}}, \bibinfo {author} {\bibfnamefont {F.}~\bibnamefont {Steglich}}, \bibinfo {author} {\bibfnamefont {M.}~\bibnamefont {Shi}}, \bibinfo {author} {\bibfnamefont {G.}~\bibnamefont {Zwicknagi}}, \bibinfo {author} {\bibfnamefont {C.}~\bibnamefont {Cao}}, \bibinfo {author}
  {\bibfnamefont {H.}~\bibnamefont {Yuan}},\ and\ \bibinfo {author} {\bibfnamefont {Y.}~\bibnamefont {Liu}},\ }\href {https://arxiv.org/abs/2309.06732} {\bibfield  {journal} {\bibinfo  {journal} {arXiv:2309.06732 [cond-mat.str-el]}\ } (\bibinfo {year} {2023})}\BibitemShut {NoStop}%
\bibitem [{\citenamefont {Pfeiffer}\ \emph {et~al.}(2023{\natexlab{a}})\citenamefont {Pfeiffer}, \citenamefont {Semeniuk}, \citenamefont {Landaeta}, \citenamefont {Borth}, \citenamefont {Geibel}, \citenamefont {Nicklas}, \citenamefont {Brando}, \citenamefont {Khim},\ and\ \citenamefont {Hassinger}}]{pfeiffer2023pressure}%
  \BibitemOpen
  \bibfield  {author} {\bibinfo {author} {\bibfnamefont {M.}~\bibnamefont {Pfeiffer}}, \bibinfo {author} {\bibfnamefont {K.}~\bibnamefont {Semeniuk}}, \bibinfo {author} {\bibfnamefont {J.~F.}\ \bibnamefont {Landaeta}}, \bibinfo {author} {\bibfnamefont {R.}~\bibnamefont {Borth}}, \bibinfo {author} {\bibfnamefont {C.}~\bibnamefont {Geibel}}, \bibinfo {author} {\bibfnamefont {M.}~\bibnamefont {Nicklas}}, \bibinfo {author} {\bibfnamefont {M.}~\bibnamefont {Brando}}, \bibinfo {author} {\bibfnamefont {S.}~\bibnamefont {Khim}},\ and\ \bibinfo {author} {\bibfnamefont {E.}~\bibnamefont {Hassinger}},\ }\href {https://arxiv.org/abs/2312.09728} {\bibfield  {journal} {\bibinfo  {journal} {arXiv:2312.09728 [cond-mat.str-el]}\ } (\bibinfo {year} {2023}{\natexlab{a}})}\BibitemShut {NoStop}%
\bibitem [{\citenamefont {Pfeiffer}\ \emph {et~al.}(2023{\natexlab{b}})\citenamefont {Pfeiffer}, \citenamefont {Semeniuk}, \citenamefont {Landaeta}, \citenamefont {Nicklas}, \citenamefont {Geibel}, \citenamefont {Brando}, \citenamefont {Khim},\ and\ \citenamefont {Hassinger}}]{pfeiffer2023exposing}%
  \BibitemOpen
  \bibfield  {author} {\bibinfo {author} {\bibfnamefont {M.}~\bibnamefont {Pfeiffer}}, \bibinfo {author} {\bibfnamefont {K.}~\bibnamefont {Semeniuk}}, \bibinfo {author} {\bibfnamefont {J.~F.}\ \bibnamefont {Landaeta}}, \bibinfo {author} {\bibfnamefont {M.}~\bibnamefont {Nicklas}}, \bibinfo {author} {\bibfnamefont {C.}~\bibnamefont {Geibel}}, \bibinfo {author} {\bibfnamefont {M.}~\bibnamefont {Brando}}, \bibinfo {author} {\bibfnamefont {S.}~\bibnamefont {Khim}},\ and\ \bibinfo {author} {\bibfnamefont {E.}~\bibnamefont {Hassinger}},\ }\href {https://arxiv.org/abs/2312.09729} {\bibfield  {journal} {\bibinfo  {journal} {arXiv:2312.09729 [cond-mat.supr-con]}\ } (\bibinfo {year} {2023}{\natexlab{b}})}\BibitemShut {NoStop}%
\bibitem [{\citenamefont {Christovam}\ \emph {et~al.}(2024)\citenamefont {Christovam}, \citenamefont {Ferreira-Carvalho}, \citenamefont {Marino}, \citenamefont {Sundermann}, \citenamefont {Takegami}, \citenamefont {Melendez-Sans}, \citenamefont {Tsuei}, \citenamefont {Hu}, \citenamefont {R{\"o}{\ss}ler}, \citenamefont {Valvidares}, \citenamefont {W.~Haverkort}, \citenamefont {Liu}, \citenamefont {D.~Bauer}, \citenamefont {H.~Tjeng}, \citenamefont {Zwicknagi},\ and\ \citenamefont {Severing}}]{christovam2024spectroscopic}%
  \BibitemOpen
  \bibfield  {author} {\bibinfo {author} {\bibfnamefont {D.~S.}\ \bibnamefont {Christovam}}, \bibinfo {author} {\bibfnamefont {M.}~\bibnamefont {Ferreira-Carvalho}}, \bibinfo {author} {\bibfnamefont {A.}~\bibnamefont {Marino}}, \bibinfo {author} {\bibfnamefont {M.}~\bibnamefont {Sundermann}}, \bibinfo {author} {\bibfnamefont {D.}~\bibnamefont {Takegami}}, \bibinfo {author} {\bibfnamefont {A.}~\bibnamefont {Melendez-Sans}}, \bibinfo {author} {\bibfnamefont {K.~D.}\ \bibnamefont {Tsuei}}, \bibinfo {author} {\bibfnamefont {Z.}~\bibnamefont {Hu}}, \bibinfo {author} {\bibfnamefont {S.}~\bibnamefont {R{\"o}{\ss}ler}}, \bibinfo {author} {\bibfnamefont {M.}~\bibnamefont {Valvidares}}, \bibinfo {author} {\bibfnamefont {M.}~\bibnamefont {W.~Haverkort}}, \bibinfo {author} {\bibfnamefont {Y.}~\bibnamefont {Liu}}, \bibinfo {author} {\bibfnamefont {E.}~\bibnamefont {D.~Bauer}}, \bibinfo {author} {\bibfnamefont {L.}~\bibnamefont {H.~Tjeng}}, \bibinfo {author} {\bibfnamefont {G.}~\bibnamefont {Zwicknagi}},\ and\ \bibinfo
  {author} {\bibfnamefont {A.}~\bibnamefont {Severing}},\ }\href {https://journals.aps.org/prl/abstract/10.1103/PhysRevLett.132.046401} {\bibfield  {journal} {\bibinfo  {journal} {Phys. Rev. Lett.}\ }\textbf {\bibinfo {volume} {132}},\ \bibinfo {pages} {046401} (\bibinfo {year} {2024})}\BibitemShut {NoStop}%
\bibitem [{\citenamefont {Chen}\ \emph {et~al.}(2024)\citenamefont {Chen}, \citenamefont {Liu}, \citenamefont {Wu}, \citenamefont {Zhang}, \citenamefont {Ye}, \citenamefont {Zhao}, \citenamefont {Song}, \citenamefont {Tian}, \citenamefont {Tan}, \citenamefont {Liu}, \citenamefont {Ye}, \citenamefont {Chen}, \citenamefont {Huang}, \citenamefont {Shen}, \citenamefont {Yuan}, \citenamefont {He}, \citenamefont {Duan},\ and\ \citenamefont {Meng}}]{chen2024exploring}%
  \BibitemOpen
  \bibfield  {author} {\bibinfo {author} {\bibfnamefont {B.}~\bibnamefont {Chen}}, \bibinfo {author} {\bibfnamefont {H.}~\bibnamefont {Liu}}, \bibinfo {author} {\bibfnamefont {Q.-Y.}\ \bibnamefont {Wu}}, \bibinfo {author} {\bibfnamefont {C.}~\bibnamefont {Zhang}}, \bibinfo {author} {\bibfnamefont {X.-Q.}\ \bibnamefont {Ye}}, \bibinfo {author} {\bibfnamefont {Y.-Z.}\ \bibnamefont {Zhao}}, \bibinfo {author} {\bibfnamefont {J.-J.}\ \bibnamefont {Song}}, \bibinfo {author} {\bibfnamefont {X.-Y.}\ \bibnamefont {Tian}}, \bibinfo {author} {\bibfnamefont {B.-L.}\ \bibnamefont {Tan}}, \bibinfo {author} {\bibfnamefont {Z.-T.}\ \bibnamefont {Liu}}, \bibinfo {author} {\bibfnamefont {M.}~\bibnamefont {Ye}}, \bibinfo {author} {\bibfnamefont {Z.-H.}\ \bibnamefont {Chen}}, \bibinfo {author} {\bibfnamefont {Y.-B.}\ \bibnamefont {Huang}}, \bibinfo {author} {\bibfnamefont {D.-W.}\ \bibnamefont {Shen}}, \bibinfo {author} {\bibfnamefont {Y.-H.}\ \bibnamefont {Yuan}}, \bibinfo {author} {\bibfnamefont {J.}~\bibnamefont {He}},
  \bibinfo {author} {\bibfnamefont {Y.-X.}\ \bibnamefont {Duan}},\ and\ \bibinfo {author} {\bibfnamefont {J.-Q.}\ \bibnamefont {Meng}},\ }\href {https://arxiv.org/abs/2403.14057} {\bibfield  {journal} {\bibinfo  {journal} {arXiv:2403.14057 [cond-mat.supr-con]}\ } (\bibinfo {year} {2024})}\BibitemShut {NoStop}%
\bibitem [{\citenamefont {Chajewski}\ and\ \citenamefont {Kaczorowski}(2024)}]{chajewski2024discovery}%
  \BibitemOpen
  \bibfield  {author} {\bibinfo {author} {\bibfnamefont {G.}~\bibnamefont {Chajewski}}\ and\ \bibinfo {author} {\bibfnamefont {D.}~\bibnamefont {Kaczorowski}},\ }\href {https://journals.aps.org/prl/abstract/10.1103/PhysRevLett.132.076504} {\bibfield  {journal} {\bibinfo  {journal} {Phys. Rev. Lett.}\ }\textbf {\bibinfo {volume} {132}},\ \bibinfo {pages} {076504} (\bibinfo {year} {2024})}\BibitemShut {NoStop}%
\bibitem [{\citenamefont {Hafner}\ \emph {et~al.}(2022)\citenamefont {Hafner}, \citenamefont {Khanenko}, \citenamefont {Eljaouhari}, \citenamefont {K{\"u}chler}, \citenamefont {Banda}, \citenamefont {Bannor}, \citenamefont {L{\"u}hmann}, \citenamefont {Landaeta}, \citenamefont {Mishra}, \citenamefont {Sheikin}, \citenamefont {Hassinger}, \citenamefont {Khim}, \citenamefont {Geibel}, \citenamefont {Zwicknagi},\ and\ \citenamefont {Brando}}]{hafner2022possible}%
  \BibitemOpen
  \bibfield  {author} {\bibinfo {author} {\bibfnamefont {D.}~\bibnamefont {Hafner}}, \bibinfo {author} {\bibfnamefont {P.}~\bibnamefont {Khanenko}}, \bibinfo {author} {\bibfnamefont {E.-O.}\ \bibnamefont {Eljaouhari}}, \bibinfo {author} {\bibfnamefont {R.}~\bibnamefont {K{\"u}chler}}, \bibinfo {author} {\bibfnamefont {J.}~\bibnamefont {Banda}}, \bibinfo {author} {\bibfnamefont {N.}~\bibnamefont {Bannor}}, \bibinfo {author} {\bibfnamefont {T.}~\bibnamefont {L{\"u}hmann}}, \bibinfo {author} {\bibfnamefont {J.~F.}\ \bibnamefont {Landaeta}}, \bibinfo {author} {\bibfnamefont {S.}~\bibnamefont {Mishra}}, \bibinfo {author} {\bibfnamefont {I.}~\bibnamefont {Sheikin}}, \bibinfo {author} {\bibfnamefont {E.}~\bibnamefont {Hassinger}}, \bibinfo {author} {\bibfnamefont {S.}~\bibnamefont {Khim}}, \bibinfo {author} {\bibfnamefont {C.}~\bibnamefont {Geibel}}, \bibinfo {author} {\bibfnamefont {G.}~\bibnamefont {Zwicknagi}},\ and\ \bibinfo {author} {\bibfnamefont {M.}~\bibnamefont {Brando}},\ }\href
  {https://journals.aps.org/prx/abstract/10.1103/PhysRevX.12.011023} {\bibfield  {journal} {\bibinfo  {journal} {Phys. Rev. X}\ }\textbf {\bibinfo {volume} {12}},\ \bibinfo {pages} {011023} (\bibinfo {year} {2022})}\BibitemShut {NoStop}%
\bibitem [{\citenamefont {Kibune}\ \emph {et~al.}(2022)\citenamefont {Kibune}, \citenamefont {Kitagawa}, \citenamefont {Kinjo}, \citenamefont {Ogata}, \citenamefont {Manago}, \citenamefont {Taniguchi}, \citenamefont {Ishida}, \citenamefont {Brando}, \citenamefont {Hassinger}, \citenamefont {Rosner}, \citenamefont {Geibel},\ and\ \citenamefont {Khim}}]{kibune2022observation}%
  \BibitemOpen
  \bibfield  {author} {\bibinfo {author} {\bibfnamefont {M.}~\bibnamefont {Kibune}}, \bibinfo {author} {\bibfnamefont {S.}~\bibnamefont {Kitagawa}}, \bibinfo {author} {\bibfnamefont {K.}~\bibnamefont {Kinjo}}, \bibinfo {author} {\bibfnamefont {S.}~\bibnamefont {Ogata}}, \bibinfo {author} {\bibfnamefont {M.}~\bibnamefont {Manago}}, \bibinfo {author} {\bibfnamefont {T.}~\bibnamefont {Taniguchi}}, \bibinfo {author} {\bibfnamefont {K.}~\bibnamefont {Ishida}}, \bibinfo {author} {\bibfnamefont {M.}~\bibnamefont {Brando}}, \bibinfo {author} {\bibfnamefont {E.}~\bibnamefont {Hassinger}}, \bibinfo {author} {\bibfnamefont {H.}~\bibnamefont {Rosner}}, \bibinfo {author} {\bibfnamefont {C.}~\bibnamefont {Geibel}},\ and\ \bibinfo {author} {\bibfnamefont {S.}~\bibnamefont {Khim}},\ }\href {https://journals.aps.org/prl/abstract/10.1103/PhysRevLett.128.057002} {\bibfield  {journal} {\bibinfo  {journal} {Phys. Rev. Lett.}\ }\textbf {\bibinfo {volume} {128}},\ \bibinfo {pages} {057002} (\bibinfo {year} {2022})}\BibitemShut
  {NoStop}%
\bibitem [{\citenamefont {Kitagawa}\ \emph {et~al.}(2022)\citenamefont {Kitagawa}, \citenamefont {Kibune}, \citenamefont {Kinjo}, \citenamefont {Manago}, \citenamefont {Taniguchi}, \citenamefont {Ishida}, \citenamefont {Brando}, \citenamefont {Hassinger}, \citenamefont {Geibel},\ and\ \citenamefont {Khim}}]{kitagawa2022two}%
  \BibitemOpen
  \bibfield  {author} {\bibinfo {author} {\bibfnamefont {S.}~\bibnamefont {Kitagawa}}, \bibinfo {author} {\bibfnamefont {M.}~\bibnamefont {Kibune}}, \bibinfo {author} {\bibfnamefont {K.}~\bibnamefont {Kinjo}}, \bibinfo {author} {\bibfnamefont {M.}~\bibnamefont {Manago}}, \bibinfo {author} {\bibfnamefont {T.}~\bibnamefont {Taniguchi}}, \bibinfo {author} {\bibfnamefont {K.}~\bibnamefont {Ishida}}, \bibinfo {author} {\bibfnamefont {M.}~\bibnamefont {Brando}}, \bibinfo {author} {\bibfnamefont {E.}~\bibnamefont {Hassinger}}, \bibinfo {author} {\bibfnamefont {C.}~\bibnamefont {Geibel}},\ and\ \bibinfo {author} {\bibfnamefont {S.}~\bibnamefont {Khim}},\ }\href {https://journals.jps.jp/doi/full/10.7566/JPSJ.91.043702} {\bibfield  {journal} {\bibinfo  {journal} {J. Phys. Soc. Jpn.}\ }\textbf {\bibinfo {volume} {91}},\ \bibinfo {pages} {043702} (\bibinfo {year} {2022})}\BibitemShut {NoStop}%
\bibitem [{\citenamefont {Ogata}\ \emph {et~al.}(2023{\natexlab{a}})\citenamefont {Ogata}, \citenamefont {Kitagawa}, \citenamefont {Kinjo}, \citenamefont {Ishida}, \citenamefont {Brando}, \citenamefont {Hassinger}, \citenamefont {Geibel},\ and\ \citenamefont {Khim}}]{ogata2023parity}%
  \BibitemOpen
  \bibfield  {author} {\bibinfo {author} {\bibfnamefont {S.}~\bibnamefont {Ogata}}, \bibinfo {author} {\bibfnamefont {S.}~\bibnamefont {Kitagawa}}, \bibinfo {author} {\bibfnamefont {K.}~\bibnamefont {Kinjo}}, \bibinfo {author} {\bibfnamefont {K.}~\bibnamefont {Ishida}}, \bibinfo {author} {\bibfnamefont {M.}~\bibnamefont {Brando}}, \bibinfo {author} {\bibfnamefont {E.}~\bibnamefont {Hassinger}}, \bibinfo {author} {\bibfnamefont {C.}~\bibnamefont {Geibel}},\ and\ \bibinfo {author} {\bibfnamefont {S.}~\bibnamefont {Khim}},\ }\href {https://journals.aps.org/prl/abstract/10.1103/PhysRevLett.130.166001} {\bibfield  {journal} {\bibinfo  {journal} {Phys. Rev. Lett.}\ }\textbf {\bibinfo {volume} {130}},\ \bibinfo {pages} {166001} (\bibinfo {year} {2023}{\natexlab{a}})}\BibitemShut {NoStop}%
\bibitem [{\citenamefont {Ogata}\ \emph {et~al.}(2023{\natexlab{b}})\citenamefont {Ogata}, \citenamefont {Kitagawa}, \citenamefont {Kibune}, \citenamefont {Ishida}, \citenamefont {Kinjo}, \citenamefont {Brando}, \citenamefont {Geibel}, \citenamefont {Khim},\ and\ \citenamefont {Hassinger}}]{ogata2023investigation}%
  \BibitemOpen
  \bibfield  {author} {\bibinfo {author} {\bibfnamefont {S.}~\bibnamefont {Ogata}}, \bibinfo {author} {\bibfnamefont {S.}~\bibnamefont {Kitagawa}}, \bibinfo {author} {\bibfnamefont {M.}~\bibnamefont {Kibune}}, \bibinfo {author} {\bibfnamefont {K.}~\bibnamefont {Ishida}}, \bibinfo {author} {\bibfnamefont {K.}~\bibnamefont {Kinjo}}, \bibinfo {author} {\bibfnamefont {M.}~\bibnamefont {Brando}}, \bibinfo {author} {\bibfnamefont {C.}~\bibnamefont {Geibel}}, \bibinfo {author} {\bibfnamefont {S.}~\bibnamefont {Khim}},\ and\ \bibinfo {author} {\bibfnamefont {E.}~\bibnamefont {Hassinger}},\ }\href {https://www.npsm-kps.org/journal/view.html?volume=73&number=12&spage=1115&year=2023} {\bibfield  {journal} {\bibinfo  {journal} {New Phys.: Sae Mulli}\ }\textbf {\bibinfo {volume} {73}},\ \bibinfo {pages} {1115} (\bibinfo {year} {2023}{\natexlab{b}})}\BibitemShut {NoStop}%
\bibitem [{\citenamefont {M{\"o}ckli}\ \emph {et~al.}(2018)\citenamefont {M{\"o}ckli}, \citenamefont {Yanase},\ and\ \citenamefont {Sigrist}}]{mockli2018orbitally}%
  \BibitemOpen
  \bibfield  {author} {\bibinfo {author} {\bibfnamefont {D.}~\bibnamefont {M{\"o}ckli}}, \bibinfo {author} {\bibfnamefont {Y.}~\bibnamefont {Yanase}},\ and\ \bibinfo {author} {\bibfnamefont {M.}~\bibnamefont {Sigrist}},\ }\href {https://journals.aps.org/prb/abstract/10.1103/PhysRevB.97.144508} {\bibfield  {journal} {\bibinfo  {journal} {Phys. Rev. B}\ }\textbf {\bibinfo {volume} {97}},\ \bibinfo {pages} {144508} (\bibinfo {year} {2018})}\BibitemShut {NoStop}%
\bibitem [{\citenamefont {Schertenleib}\ \emph {et~al.}(2021)\citenamefont {Schertenleib}, \citenamefont {Fischer},\ and\ \citenamefont {Sigrist}}]{schertenleib2021unusual}%
  \BibitemOpen
  \bibfield  {author} {\bibinfo {author} {\bibfnamefont {E.~G.}\ \bibnamefont {Schertenleib}}, \bibinfo {author} {\bibfnamefont {M.~H.}\ \bibnamefont {Fischer}},\ and\ \bibinfo {author} {\bibfnamefont {M.}~\bibnamefont {Sigrist}},\ }\href {https://journals.aps.org/prresearch/cited-by/10.1103/PhysRevResearch.3.023179} {\bibfield  {journal} {\bibinfo  {journal} {Phys. Rev. Research}\ }\textbf {\bibinfo {volume} {3}},\ \bibinfo {pages} {023179} (\bibinfo {year} {2021})}\BibitemShut {NoStop}%
\bibitem [{\citenamefont {Kaur}\ \emph {et~al.}(2005)\citenamefont {Kaur}, \citenamefont {Agterberg},\ and\ \citenamefont {Sigrist}}]{kaur2005helical}%
  \BibitemOpen
  \bibfield  {author} {\bibinfo {author} {\bibfnamefont {R.}~\bibnamefont {Kaur}}, \bibinfo {author} {\bibfnamefont {D.}~\bibnamefont {Agterberg}},\ and\ \bibinfo {author} {\bibfnamefont {M.}~\bibnamefont {Sigrist}},\ }\href {https://journals.aps.org/prl/abstract/10.1103/PhysRevLett.94.137002} {\bibfield  {journal} {\bibinfo  {journal} {Phys. Rev. Lett.}\ }\textbf {\bibinfo {volume} {94}},\ \bibinfo {pages} {137002} (\bibinfo {year} {2005})}\BibitemShut {NoStop}%
\bibitem [{\citenamefont {Oka}\ \emph {et~al.}(2006)\citenamefont {Oka}, \citenamefont {Ichioka},\ and\ \citenamefont {Machida}}]{oka2006surface}%
  \BibitemOpen
  \bibfield  {author} {\bibinfo {author} {\bibfnamefont {M.}~\bibnamefont {Oka}}, \bibinfo {author} {\bibfnamefont {M.}~\bibnamefont {Ichioka}},\ and\ \bibinfo {author} {\bibfnamefont {K.}~\bibnamefont {Machida}},\ }\href {https://journals.aps.org/prb/abstract/10.1103/PhysRevB.73.214509} {\bibfield  {journal} {\bibinfo  {journal} {Phys. Rev. B}\ }\textbf {\bibinfo {volume} {73}},\ \bibinfo {pages} {214509} (\bibinfo {year} {2006})}\BibitemShut {NoStop}%
\bibitem [{\citenamefont {Dimitrova}\ and\ \citenamefont {Feigel’Man}(2007)}]{dimitrova2007theory}%
  \BibitemOpen
  \bibfield  {author} {\bibinfo {author} {\bibfnamefont {O.}~\bibnamefont {Dimitrova}}\ and\ \bibinfo {author} {\bibfnamefont {M.}~\bibnamefont {Feigel’Man}},\ }\href {https://journals.aps.org/prb/abstract/10.1103/PhysRevB.76.014522} {\bibfield  {journal} {\bibinfo  {journal} {Phys. Rev. B}\ }\textbf {\bibinfo {volume} {76}},\ \bibinfo {pages} {014522} (\bibinfo {year} {2007})}\BibitemShut {NoStop}%
\bibitem [{\citenamefont {Hiasa}\ and\ \citenamefont {Ikeda}(2008)}]{hiasa2008orbital}%
  \BibitemOpen
  \bibfield  {author} {\bibinfo {author} {\bibfnamefont {N.}~\bibnamefont {Hiasa}}\ and\ \bibinfo {author} {\bibfnamefont {R.}~\bibnamefont {Ikeda}},\ }\href {https://journals.aps.org/prb/abstract/10.1103/PhysRevB.78.224514} {\bibfield  {journal} {\bibinfo  {journal} {Phys. Rev. B}\ }\textbf {\bibinfo {volume} {78}},\ \bibinfo {pages} {224514} (\bibinfo {year} {2008})}\BibitemShut {NoStop}%
\bibitem [{\citenamefont {Iniotakis}\ \emph {et~al.}(2008)\citenamefont {Iniotakis}, \citenamefont {Fujimoto},\ and\ \citenamefont {Sigrist}}]{iniotakis2008fractional}%
  \BibitemOpen
  \bibfield  {author} {\bibinfo {author} {\bibfnamefont {C.}~\bibnamefont {Iniotakis}}, \bibinfo {author} {\bibfnamefont {S.}~\bibnamefont {Fujimoto}},\ and\ \bibinfo {author} {\bibfnamefont {M.}~\bibnamefont {Sigrist}},\ }\href {https://journals.jps.jp/doi/10.1143/JPSJ.77.083701} {\bibfield  {journal} {\bibinfo  {journal} {J. Phys. Soc. Jpn.}\ }\textbf {\bibinfo {volume} {77}},\ \bibinfo {pages} {083701} (\bibinfo {year} {2008})}\BibitemShut {NoStop}%
\bibitem [{\citenamefont {Matsunaga}\ \emph {et~al.}(2008)\citenamefont {Matsunaga}, \citenamefont {Hiasa},\ and\ \citenamefont {Ikeda}}]{matsunaga2008modulated}%
  \BibitemOpen
  \bibfield  {author} {\bibinfo {author} {\bibfnamefont {Y.}~\bibnamefont {Matsunaga}}, \bibinfo {author} {\bibfnamefont {N.}~\bibnamefont {Hiasa}},\ and\ \bibinfo {author} {\bibfnamefont {R.}~\bibnamefont {Ikeda}},\ }\href {https://journals.aps.org/prb/abstract/10.1103/PhysRevB.78.220508} {\bibfield  {journal} {\bibinfo  {journal} {Phys. Rev. B}\ }\textbf {\bibinfo {volume} {78}},\ \bibinfo {pages} {220508} (\bibinfo {year} {2008})}\BibitemShut {NoStop}%
\bibitem [{\citenamefont {Hiasa}\ \emph {et~al.}(2009)\citenamefont {Hiasa}, \citenamefont {Saiki},\ and\ \citenamefont {Ikeda}}]{hiasa2009vortex}%
  \BibitemOpen
  \bibfield  {author} {\bibinfo {author} {\bibfnamefont {N.}~\bibnamefont {Hiasa}}, \bibinfo {author} {\bibfnamefont {T.}~\bibnamefont {Saiki}},\ and\ \bibinfo {author} {\bibfnamefont {R.}~\bibnamefont {Ikeda}},\ }\href {https://journals.aps.org/prb/abstract/10.1103/PhysRevB.80.014501} {\bibfield  {journal} {\bibinfo  {journal} {Phys. Rev. B}\ }\textbf {\bibinfo {volume} {80}},\ \bibinfo {pages} {014501} (\bibinfo {year} {2009})}\BibitemShut {NoStop}%
\bibitem [{\citenamefont {Kashyap}\ and\ \citenamefont {Agterberg}(2013)}]{kashyap2013vortices}%
  \BibitemOpen
  \bibfield  {author} {\bibinfo {author} {\bibfnamefont {M.}~\bibnamefont {Kashyap}}\ and\ \bibinfo {author} {\bibfnamefont {D.}~\bibnamefont {Agterberg}},\ }\href {https://journals.aps.org/prb/abstract/10.1103/PhysRevB.88.104515#:~:text=The%20absence%20of%20inversion%20symmetry%20in%20noncentrosymmetric%20superconductors,%7B%7D_%20%7B3%7D%24B%20or%20Mo%24%20%7B%7D_%20%7B3%7D%24Al%24%20%7B%7D_%20%7B2%7D%24C.} {\bibfield  {journal} {\bibinfo  {journal} {Phys. Rev. B}\ }\textbf {\bibinfo {volume} {88}},\ \bibinfo {pages} {104515} (\bibinfo {year} {2013})}\BibitemShut {NoStop}%
\bibitem [{\citenamefont {Nikoli{\'c}}(2014)}]{nikolic2014vortices}%
  \BibitemOpen
  \bibfield  {author} {\bibinfo {author} {\bibfnamefont {P.}~\bibnamefont {Nikoli{\'c}}},\ }\href {https://journals.aps.org/pra/abstract/10.1103/PhysRevA.90.023623} {\bibfield  {journal} {\bibinfo  {journal} {Phys. Rev. A}\ }\textbf {\bibinfo {volume} {90}},\ \bibinfo {pages} {023623} (\bibinfo {year} {2014})}\BibitemShut {NoStop}%
\bibitem [{\citenamefont {Dan}\ and\ \citenamefont {Ikeda}(2015)}]{dan2015quasiclassical}%
  \BibitemOpen
  \bibfield  {author} {\bibinfo {author} {\bibfnamefont {Y.}~\bibnamefont {Dan}}\ and\ \bibinfo {author} {\bibfnamefont {R.}~\bibnamefont {Ikeda}},\ }\href {https://journals.aps.org/prb/abstract/10.1103/PhysRevB.92.144504} {\bibfield  {journal} {\bibinfo  {journal} {Phys. Rev. B}\ }\textbf {\bibinfo {volume} {92}},\ \bibinfo {pages} {144504} (\bibinfo {year} {2015})}\BibitemShut {NoStop}%
\bibitem [{\citenamefont {Cavanagh}\ \emph {et~al.}(2022)\citenamefont {Cavanagh}, \citenamefont {Shishidou}, \citenamefont {Weinert}, \citenamefont {Brydon},\ and\ \citenamefont {Agterberg}}]{cavanagh2022nonsymmorphic}%
  \BibitemOpen
  \bibfield  {author} {\bibinfo {author} {\bibfnamefont {D.}~\bibnamefont {Cavanagh}}, \bibinfo {author} {\bibfnamefont {T.}~\bibnamefont {Shishidou}}, \bibinfo {author} {\bibfnamefont {M.}~\bibnamefont {Weinert}}, \bibinfo {author} {\bibfnamefont {P.}~\bibnamefont {Brydon}},\ and\ \bibinfo {author} {\bibfnamefont {D.~F.}\ \bibnamefont {Agterberg}},\ }\href {https://journals.aps.org/prb/abstract/10.1103/PhysRevB.105.L020505} {\bibfield  {journal} {\bibinfo  {journal} {Phys. Rev. B}\ }\textbf {\bibinfo {volume} {105}},\ \bibinfo {pages} {L020505} (\bibinfo {year} {2022})}\BibitemShut {NoStop}%
\bibitem [{\citenamefont {Hubbard}(1959)}]{hubbard1959calculation}%
  \BibitemOpen
  \bibfield  {author} {\bibinfo {author} {\bibfnamefont {J.}~\bibnamefont {Hubbard}},\ }\href {https://journals.aps.org/prl/abstract/10.1103/PhysRevLett.3.77} {\bibfield  {journal} {\bibinfo  {journal} {Phys. Rev. Lett.}\ }\textbf {\bibinfo {volume} {3}},\ \bibinfo {pages} {77} (\bibinfo {year} {1959})}\BibitemShut {NoStop}%
\bibitem [{\citenamefont {Adachi}\ and\ \citenamefont {Ikeda}(2003)}]{adachi2003effects}%
  \BibitemOpen
  \bibfield  {author} {\bibinfo {author} {\bibfnamefont {H.}~\bibnamefont {Adachi}}\ and\ \bibinfo {author} {\bibfnamefont {R.}~\bibnamefont {Ikeda}},\ }\href {https://journals.aps.org/prb/abstract/10.1103/PhysRevB.68.184510} {\bibfield  {journal} {\bibinfo  {journal} {Phys. Rev. B}\ }\textbf {\bibinfo {volume} {68}},\ \bibinfo {pages} {184510} (\bibinfo {year} {2003})}\BibitemShut {NoStop}%
\bibitem [{\citenamefont {Adachi}\ and\ \citenamefont {Ikeda}(2015)}]{adachi2015possible}%
  \BibitemOpen
  \bibfield  {author} {\bibinfo {author} {\bibfnamefont {K.}~\bibnamefont {Adachi}}\ and\ \bibinfo {author} {\bibfnamefont {R.}~\bibnamefont {Ikeda}},\ }\href {https://journals.jps.jp/doi/abs/10.7566/JPSJ.84.064712} {\bibfield  {journal} {\bibinfo  {journal} {J. Phys. Soc. Jpn.}\ }\textbf {\bibinfo {volume} {84}},\ \bibinfo {pages} {064712} (\bibinfo {year} {2015})}\BibitemShut {NoStop}%
\bibitem [{sup()}]{supplemental}%
  \BibitemOpen
  \href@noop {} {}\bibinfo {note} {See Supplemental Materials for details of the derivation of GL free energy functional.}\BibitemShut {Stop}%
\bibitem [{\citenamefont {Maruyama}\ \emph {et~al.}(2012)\citenamefont {Maruyama}, \citenamefont {Sigrist},\ and\ \citenamefont {Yanase}}]{maruyama2012locally}%
  \BibitemOpen
  \bibfield  {author} {\bibinfo {author} {\bibfnamefont {D.}~\bibnamefont {Maruyama}}, \bibinfo {author} {\bibfnamefont {M.}~\bibnamefont {Sigrist}},\ and\ \bibinfo {author} {\bibfnamefont {Y.}~\bibnamefont {Yanase}},\ }\href {https://journals.jps.jp/doi/abs/10.1143/JPSJ.81.034702?journalCode=jpsj} {\bibfield  {journal} {\bibinfo  {journal} {J. Phys. Soc. Jpn.}\ }\textbf {\bibinfo {volume} {81}},\ \bibinfo {pages} {034702} (\bibinfo {year} {2012})}\BibitemShut {NoStop}%
\bibitem [{\citenamefont {Maki}(1966)}]{maki1966effect}%
  \BibitemOpen
  \bibfield  {author} {\bibinfo {author} {\bibfnamefont {K.}~\bibnamefont {Maki}},\ }\href {https://journals.aps.org/pr/abstract/10.1103/PhysRev.148.362} {\bibfield  {journal} {\bibinfo  {journal} {Phys. Rev.}\ }\textbf {\bibinfo {volume} {148}},\ \bibinfo {pages} {362} (\bibinfo {year} {1966})}\BibitemShut {NoStop}%
\bibitem [{\citenamefont {Nica}\ \emph {et~al.}(2015)\citenamefont {Nica}, \citenamefont {Yu},\ and\ \citenamefont {Si}}]{nica2015glide}%
  \BibitemOpen
  \bibfield  {author} {\bibinfo {author} {\bibfnamefont {E.~M.}\ \bibnamefont {Nica}}, \bibinfo {author} {\bibfnamefont {R.}~\bibnamefont {Yu}},\ and\ \bibinfo {author} {\bibfnamefont {Q.}~\bibnamefont {Si}},\ }\href {https://journals.aps.org/prb/abstract/10.1103/PhysRevB.92.174520} {\bibfield  {journal} {\bibinfo  {journal} {Phys. Rev. B}\ }\textbf {\bibinfo {volume} {92}},\ \bibinfo {pages} {174520} (\bibinfo {year} {2015})}\BibitemShut {NoStop}%
\bibitem [{\citenamefont {Matsuda}\ and\ \citenamefont {Shimahara}(2007)}]{matsuda2007fulde}%
  \BibitemOpen
  \bibfield  {author} {\bibinfo {author} {\bibfnamefont {Y.}~\bibnamefont {Matsuda}}\ and\ \bibinfo {author} {\bibfnamefont {H.}~\bibnamefont {Shimahara}},\ }\href {https://journals.jps.jp/doi/10.1143/JPSJ.76.051005} {\bibfield  {journal} {\bibinfo  {journal} {J. Phys. Soc. Jpn.}\ }\textbf {\bibinfo {volume} {76}},\ \bibinfo {pages} {051005} (\bibinfo {year} {2007})}\BibitemShut {NoStop}%
\bibitem [{\citenamefont {Kasamatsu}\ \emph {et~al.}(2005)\citenamefont {Kasamatsu}, \citenamefont {Tsubota},\ and\ \citenamefont {Ueda}}]{kasamatsu2005spin}%
  \BibitemOpen
  \bibfield  {author} {\bibinfo {author} {\bibfnamefont {K.}~\bibnamefont {Kasamatsu}}, \bibinfo {author} {\bibfnamefont {M.}~\bibnamefont {Tsubota}},\ and\ \bibinfo {author} {\bibfnamefont {M.}~\bibnamefont {Ueda}},\ }\href {https://journals.aps.org/pra/abstract/10.1103/PhysRevA.71.043611} {\bibfield  {journal} {\bibinfo  {journal} {Phys. Rev. A}\ }\textbf {\bibinfo {volume} {71}},\ \bibinfo {pages} {043611} (\bibinfo {year} {2005})}\BibitemShut {NoStop}%
\bibitem [{\citenamefont {G{\"o}bel}\ \emph {et~al.}(2021)\citenamefont {G{\"o}bel}, \citenamefont {Mertig},\ and\ \citenamefont {Tretiakov}}]{gobel2021beyond}%
  \BibitemOpen
  \bibfield  {author} {\bibinfo {author} {\bibfnamefont {B.}~\bibnamefont {G{\"o}bel}}, \bibinfo {author} {\bibfnamefont {I.}~\bibnamefont {Mertig}},\ and\ \bibinfo {author} {\bibfnamefont {O.~A.}\ \bibnamefont {Tretiakov}},\ }\href {https://www.sciencedirect.com/science/article/pii/S0370157320303525} {\bibfield  {journal} {\bibinfo  {journal} {Phys. Rep.}\ }\textbf {\bibinfo {volume} {895}},\ \bibinfo {pages} {1} (\bibinfo {year} {2021})}\BibitemShut {NoStop}%
\bibitem [{\citenamefont {Salomaa}\ and\ \citenamefont {Volovik}(1987)}]{Salomaa-Volovik}%
  \BibitemOpen
  \bibfield  {author} {\bibinfo {author} {\bibfnamefont {M.~M.}\ \bibnamefont {Salomaa}}\ and\ \bibinfo {author} {\bibfnamefont {G.~E.}\ \bibnamefont {Volovik}},\ }\href {https://doi.org/10.1103/RevModPhys.59.533} {\bibfield  {journal} {\bibinfo  {journal} {Rev. Mod. Phys.}\ }\textbf {\bibinfo {volume} {59}},\ \bibinfo {pages} {533} (\bibinfo {year} {1987})}\BibitemShut {NoStop}%
\bibitem [{\citenamefont {Tokiwa}\ \emph {et~al.}(2023)\citenamefont {Tokiwa}, \citenamefont {Sakai}, \citenamefont {Kambe}, \citenamefont {Opletal}, \citenamefont {Yamamoto}, \citenamefont {Kimata}, \citenamefont {Awaji}, \citenamefont {Sasaki}, \citenamefont {Yanase}, \citenamefont {Haga},\ and\ \citenamefont {Tokunaga}}]{Tokiwa2023}%
  \BibitemOpen
  \bibfield  {author} {\bibinfo {author} {\bibfnamefont {Y.}~\bibnamefont {Tokiwa}}, \bibinfo {author} {\bibfnamefont {H.}~\bibnamefont {Sakai}}, \bibinfo {author} {\bibfnamefont {S.}~\bibnamefont {Kambe}}, \bibinfo {author} {\bibfnamefont {P.}~\bibnamefont {Opletal}}, \bibinfo {author} {\bibfnamefont {E.}~\bibnamefont {Yamamoto}}, \bibinfo {author} {\bibfnamefont {M.}~\bibnamefont {Kimata}}, \bibinfo {author} {\bibfnamefont {S.}~\bibnamefont {Awaji}}, \bibinfo {author} {\bibfnamefont {T.}~\bibnamefont {Sasaki}}, \bibinfo {author} {\bibfnamefont {Y.}~\bibnamefont {Yanase}}, \bibinfo {author} {\bibfnamefont {Y.}~\bibnamefont {Haga}},\ and\ \bibinfo {author} {\bibfnamefont {Y.}~\bibnamefont {Tokunaga}},\ }\href {https://doi.org/10.1103/PhysRevB.108.144502} {\bibfield  {journal} {\bibinfo  {journal} {Phys. Rev. B}\ }\textbf {\bibinfo {volume} {108}},\ \bibinfo {pages} {144502} (\bibinfo {year} {2023})}\BibitemShut {NoStop}%
\end{thebibliography}%
\bibliographystyle{apsrev4-2}

\begin{widetext}
\appendix
\section{Appendix A: Derivation of Ginzburg-Landau free energy functional}
In the following, we explain how to derive the GL free energy functional based on the procedure similar to Refs.~\cite{dimitrova2007theory,adachi2003effects,hiasa2009vortex,hiasa2008orbital,adachi2015possible}.

\subsection{1.\hspace{10pt}Green function}
The Peierls phase approximation (quasiclassical approximation)~\cite{adachi2003effects} is applied to the non-interacting Green function of $\mathcal{H}_{0}$
\begin{equation}
    \label{eq:ap-1}
    \hat{G}^{(0)}(\bm{r},\bm{r}',i\omega_n)
    =
    \hat{G}^{'(0)}(\bm{r}-\bm{r}',i\omega_n)
    e^{-ie\int^{\bm{r}}_{\bm{r}'}d\bm{s}\cdot\bm{A}},
    \tag{A1}
\end{equation}
where $\omega_n=(2n+1)\pi T$ is the fermionic Matsubara frequency.
This approximation is valid if $k_{\rm F} r_H\gg 1$, where $k_{\rm F}$ is the Fermi momentum, and $r_H=(2eH)^{-1/2}$ is the magnetic length.
$\hat{G}^{'(0)}$ is the Green function of $\mathcal{H}_{0}^{\bm{A}=0}$, in which the orbital effect is neglected.
$\mathcal{H}_{0}^{\bm{A}=0}$ is diagonalized by an appropriate unitary transformation with $\hat{U}(\bm{k})$ and the resulting band structure consists of two almost degenerated bands slightly split by the Zeeman energy~\cite{lee2023linear},
\begin{equation}
    \label{eq:ap-2}
    E_{\nu,\pm}(\bm{k})=\xi(\bm{k})+(-1)^{\nu}\sqrt{(t_\perp\pm h')^2+\alpha^2|\bm{g}(\bm{k})|^2},
    \tag{A2}
\end{equation}
where $\nu=1,2$, and $h'=\mu_{\rm B} H$.
Then, the non-interacting Green functions in the band basis $G_{\nu,\lambda}^{'(0)}(\bm{r}-\bm{r}',i\omega_n)$ are given as the Fourier transform of $G_{\nu,\lambda}^{'(0)}(\bm{k},i\omega_n)=(i\omega_n-E_{\nu,\lambda}(\bm{k}))^{-1}$.

\subsection{2.\hspace{10pt}Quadratic term}
The order parameter on each layer $\Delta_m (m=1,2)$ is decomposed into the sublattice-symmetric component $\Delta_e$ and sublattice-antisymmetric component $\Delta_o$.
\begin{align}
    \label{eq:ap-3}
    \Delta_e(\bm{r})=&\frac{1}{2}(\Delta_1(\bm{r})+\Delta_2(\bm{r})),
    \tag{A3}\\
        \label{eq:ap-4}
    \Delta_o(\bm{r})=&\frac{1}{2}(\Delta_1(\bm{r})-\Delta_2(\bm{r})).
    \tag{A4}
\end{align}
Using the functional integral method, we obtain the differential operators $\hat{\mathcal{K}}_{j}^{(2)}(\bm{\Pi})\ (j=e,o)$ in the quadratic term of the GL free energy functional
\begin{equation}
    \label{eq:ap-5}
    \mathcal{F}^{(2)}
    =
    \sum_{j}\int d^2\bm{r}
    \ \Delta^*_j(\bm{r})\left(\frac{2}{V_j}-\hat{\mathcal{K}}_{j}^{(2)}(\bm{\Pi})\right)\Delta_j(\bm{r}),
    \tag{A5}
\end{equation}
as
\begin{align}
    \hat{\mathcal{K}}_{e}^{(2)}(\bm{\Pi})
    =&
    \frac{T}{\Omega}
    \sum_{\bm{k},\omega_n,\nu}
    G^{'(0)}_{\nu,+}(\bm{k},i\omega_n)
    G^{'(0)}_{\nu,-}(\bm{\Pi}-\bm{k},-i\omega_n)
    \notag\\
    \label{eq:ap-6}
    =&
    2\pi TN(0)
    \int^{\infty}_{\rho_c}
    \frac{d\rho}{\sinh(2\pi T\rho)}
    \cos(2h'\rho\cos\chi)
    \left\langle
        e^{i\rho \bm{v}_{\rm F}\cdot \bm{\Pi}}
        \right\rangle_{\hat{\bm{k}}},
    \tag{A6}\\
    \hat{\mathcal{K}}_{o}^{(2)}(\bm{\Pi})
    =&
    \frac{T}{2\Omega}
    \sin^2\chi
    \sum_{\bm{k},\omega_n,\nu,\lambda}
    G^{'(0)}_{\nu,\lambda}(\bm{k},i\omega_n)
    G^{'(0)}_{\nu,\lambda}(\bm{\Pi}-\bm{k},-i\omega_n)\notag\\
    \label{eq:ap-7}
    =&
    2\pi TN(0)
    \sin^2\chi
    \int^{\infty}_{\rho_c}
    \frac{d\rho}{\sinh(2\pi T\rho)}
    \left\langle
        e^{i\rho \bm{v}_{\rm F}\cdot \bm{\Pi}}
        \right\rangle_{\hat{\bm{k}}},
    \tag{A7}
\end{align}
where $\bm{\Pi}=-i\nabla+2e\bm{A}$ is the gauge invariant differential operator and $\Omega=L_xL_y$ is the area of layers.
A (normalized) difference between the density of states on the two bands
\begin{equation}
    \label{eq:ap-8}
    \delta N\equiv\frac{|N_1(0)-N_2(0)|}{N_1(0)+N_2(0)},
    \tag{A8}
\end{equation}
is on the order of $(\sqrt{\alpha^2+t_\perp^2}/E_{\rm F})^1$~\cite{hiasa2009vortex,matsunaga2008modulated}, and we consider the limit $\delta N\to 0\ [N_1(0)=N_2(0)=N(0)]$ for simplicity.
Also, the Fermi velocity $\bm{v}_{\rm F}=v_{\rm F}\hat{\bm{k}}$ is assumed to be the same on both bands.
To perform the integral over momentum, semiclassical approximation~\cite{dimitrova2007theory}
\begin{equation}
    \label{eq:ap-9}
    \sum_{\bm{k}}\cdots
    =\frac{\Omega N(0)}{2}\int^{\infty}_{-\infty}d\xi\langle\cdots\rangle_{\hat{\bm{k}}},
    \tag{A9}
\end{equation}
is applied, where the bracket $\langle\cdots\rangle_{\hat{\bm{k}}}$ represents an angular average over the Fermi surface, $\langle\cdots\rangle_{\hat{\bm{k}}}=\int d\theta_{\hat{\bm{k}}}/(2\pi)\cdots $ with $ \theta_{\hat{\bm{k}}}=\tan^{-1}(\hat{k}_y/\hat{k}_x)$.
Note that each band is doubly degenerated.

In the presence of a magnetic field, the eigenfunctions of $\hat{\mathcal{K}}_{j}^{(2)}(\bm{\Pi})$ are basis functions of Landau levels (LLs).
$\Delta_j$ is expanded via the LLs $\{\psi_{Nq}\}_{N,q}$,
\begin{equation}
    \label{eq:ap-10}
    \Delta_j(\bm{r})
    =\sum_{N,q}\Delta_j(N,q)\psi_{Nq}(\bm{r}).
    \tag{A10}
\end{equation}
It can be shown that $e^{i\rho \bm{v}_{\rm F}\cdot \bm{\Pi}}$ acts on $\psi_{Nq}$ as
\begin{align}
    \left\langle
        e^{i\rho \bm{v}_{\rm F}\cdot \bm{\Pi}}
        \right\rangle_{\hat{\bm{k}}}
    \psi_{Nq}(\bm{r})
    =&
    e^{-\rho^2/4\tau_H^2}
    \left\langle
        e^{i\rho \hat{k}_{-}\pi_{+}/\sqrt{2}\tau_H}
        e^{i\rho \hat{k}_{+}\pi_{-}/\sqrt{2}\tau_H}
        \right\rangle_{\hat{\bm{k}}}
    \psi_{Nq}(\bm{r})\notag\\
    \label{eq:ap-11}
    =&e^{-\rho^2/4\tau_H^2}
    L_N\left(\frac{\rho^2}{2\tau_H^2}\right)
    \psi_{Nq}(\bm{r}), 
    \tag{A11}
\end{align}
by using the Baker-Campbell-Hausdorff formula, where $\hat{k}_{\pm}=\hat{k}_{x}\pm i\hat{k}_{y}$, $\tau_H=r_H/v_{\rm F}$, and
\begin{equation}
    \label{eq:ap-12}
    \pi_{\pm}=\frac{r_H}{\sqrt{2}}
    (\Pi_x\pm i\Pi_y),
    \tag{A12}
\end{equation}
are the creation and annihilation operators of LLs.
Substituting Eqs.~\eqref{eq:ap-6}, \eqref{eq:ap-7}, \eqref{eq:ap-10}, \eqref{eq:ap-11} into Eq.~\eqref{eq:ap-5}, the explicit expressions of $\hat{\mathcal{K}}_{j}^{(2)}(\bm{\Pi})$ are obtained as,
\begin{align}
    \label{eq:ap-13}
    \int d^2\bm{r}
    \ \Delta_e^*(\bm{r})\hat{\mathcal{K}}_{e}^{(2)}(\bm{\Pi})\Delta_e(\bm{r})
    =&
    2\pi TN(0)\sum_{N,q}
    |\Delta_e(N,q)|^2
    \int^{\infty}_{\rho_c}\frac{d\rho}{\sinh(2\pi T\rho)}
    \cos(2h'\rho\cos\chi)
    e^{-\rho^2/4\tau_H^2}
    L_{N}\left(\frac{\rho^2}{2\tau_H^2}\right),
    \tag{A13}\\
    \int d^2\bm{r}
    \ \Delta_o^*(\bm{r})\hat{\mathcal{K}}_{o}^{(2)}(\bm{\Pi})\Delta_o(\bm{r})
    =&
    2\pi TN(0)\sin^2\chi\sum_{N,q}
    |\Delta_o(N,q)|^2
    \int^{\infty}_{\rho_c}\frac{d\rho}{\sinh(2\pi T\rho)}
    e^{-\rho^2/4\tau_H^2}
    L_{N}\left(\frac{\rho^2}{2\tau_H^2}\right).
    \tag{A14}
\end{align}
Here, the auxiliary variable $\rho$ is introduced by using the identity
\begin{equation}
    \frac{1}{X}=\int^{\infty}_{0}d\rho\ e^{-X\rho},
    \tag{A15}
\end{equation}
where $X>0$.
As is often done, we eliminate the cutoff $\rho_c$ by introducing the zero-field transition temperature $T_{c0}^{j}$ through the relation
\begin{align}
    \frac{1}{N(0)V_e}
    =&\ln\frac{T}{T_{c0}^{e}}
    +\int^{\infty}_{\rho_c}\frac{\pi Td\rho}{\sinh(2\pi T\rho)},
    \tag{A16}\\
    \frac{1}{N(0)V_o}
    =&\ln\frac{T}{T_{c0}^{o}}
    +\sin^2\chi\int^{\infty}_{\rho_c}\frac{\pi Td\rho}{\sinh(2\pi T\rho)}.
    \tag{A17}
\end{align}
Finally, we reach the expression of $\mathcal{F}^{(2)}$,
\begin{align}
    &\mathcal{F}^{(2)}
    =2N(0)
    \sum_{N,q,j}
    E_{N}^{\,j}
    |\Delta_j(N,q)|^2,
    \tag{A18}\\
    &E_{N}^{\,e}
    =
    \ln\left(\frac{T}{T_{\mathrm{c}0}^{e}}\right)+\int^{\infty}_{0}\frac{\pi Td\rho}{\sinh(2\pi T\rho)}
    \tilde{\mathcal{K}}^{(2)}_{e}(\rho,N),
    \tag{A19}\\
    &E_{N}^{\,o}
    =
    \sin^2\chi\left[\ln\left(\frac{T}{T_{\mathrm{c}0}^{o}}\right)+\int^{\infty}_{0}\frac{\pi Td\rho}{\sinh(2\pi T\rho)}
    \tilde{\mathcal{K}}^{(2)}_{o}(\rho,N)\right],
    \tag{A20}\\
    &\tilde{\mathcal{K}}^{(2)}_{e}(\rho,N)
    =1-\cos(2h'\rho\cos\chi)
    e^{-\rho^2/4\tau_H^2}
    L_N\left(\frac{\rho^2}{2\tau_H^2}\right),
    \tag{A21}\\
    &\tilde{\mathcal{K}}^{(2)}_{o}(\rho,N)
    =1-
    e^{-\rho^2/4\tau_H^2}
    L_N\left(\frac{\rho^2}{2\tau_H^2}\right).
    \tag{A22}
\end{align}
Note that $E_{N}^{\,o}$ vanishes when $\alpha=0$.
That is, the PDW state [$(\Delta_e,\Delta_o)=(0,\Delta)$] does not appear in the phase diagram at all.
The reason why the presence of the ASOC is necessary to stabilize the PDW state is as follows. If $\alpha=0$, then the Cooper pair in the PDW state is formed by interband pairing, which is unstable~\cite{yoshida2012pair}.

\subsection{3.\hspace{10pt}Quartic term}
Next, the quartic term of the GL free energy functional takes the form
\begin{equation}
    \label{eq:ap-14}
    \mathcal{F}^{(4)}
    =\frac{1}{2}
    \sum_{\{j_i\}}\int d^2\bm{r}\ 
    \mathrm{Re}
    \left[\hat{\mathcal{K}}^{(4)}(\{\bm{\Pi}_i\},\{j_i\})
    \Delta_{j_1}^*(\bm{r}_{1})\Delta_{j_2}(\bm{r}_{2})\left.\Delta_{j_3}^*(\bm{r}_{3})
    \Delta_{j_4}(\bm{r}_{4})\right|_{\bm{r}_{i}=\bm{r}}\right],
    \tag{A23}
\end{equation}
where
\begin{align}
    \label{eq:ap-15}
    \hat{\mathcal{K}}^{(4)}(\{\bm{\Pi}_i\},\{j_i\})
    =&\frac{T}{\Omega}
    \sum_{\bm{k},\omega_{n}}
    \sum_{\{\overline{\nu}_i\}}
    C
    G^{'(0)}_{\overline{\nu}_1}(\bm{k},i\omega_n)
    G^{'(0)}_{\overline{\nu}_2}(-\bm{k}+\bm{\Pi}_1^*,-i\omega_n)\notag\\
    &\ \ \ \times G^{'(0)}_{\overline{\nu}_3}(-\bm{k}+\bm{\Pi}_2,-i\omega_n)
    G^{'(0)}_{\overline{\nu}_4}(\bm{k}-\bm{\Pi}_2+\bm{\Pi}_3^*,i\omega_n)\notag\\
    =&2\pi TN(0)\sum_{\{\overline{\nu}_i\}}
    \tilde{C}
    \int^{\infty}_{0}\frac{d\rho_1d\rho_2d\rho_3}{\sinh[2\pi T(\rho_1+\rho_2+\rho_3)]}\notag\\
    &\ \ \ \times\cos(\rho_1\alpha_1+\rho_2\alpha_2+\rho_3\alpha_3)
    \left\langle
        e^{i\bm{v}_F\cdot (\rho_1\bm{\Pi}_1^*+\rho_2\bm{\Pi}_2+\rho_3\bm{\Pi}_3^*)}
    \right\rangle_{\hat{\bm{k}}}.
    \tag{A24}
\end{align}
Here, the abbreviated notation $\overline{\nu}=(\nu,\lambda)$ is employed, and $\alpha_1=\alpha_1(\{\overline{\nu}_i\})=E_{\overline{\nu}_1}-E_{\overline{\nu}_2},\ \alpha_2=\alpha_2(\{\overline{\nu}_i\})=E_{\overline{\nu}_1}-E_{\overline{\nu}_3},\ \alpha_3=\alpha_3(\{\overline{\nu}_i\})=E_{\overline{\nu}_4}-E_{\overline{\nu}_3}$.
The coefficients $C=C(\{j_i\},\{\overline{\nu}_i\})$ and $\tilde{C}=\tilde{C}(\{j_i\},\{\overline{\nu}_i\})$ are products of the matrix elements of $\hat{U}(\bm{k})$.
The nonzero components of $\tilde{C}(\{j_i\},\{\overline{\nu}_i\})$ on the Fermi surface $\lambda=1$ are
\begin{align*}
    \tilde{C}(\{e,e,e,e\},&\{\overline{1},\overline{2},\overline{2},\overline{1}\})
    =
    \tilde{C}(\{e,e,e,e\},\{\overline{2},\overline{1},\overline{1},\overline{2}\})
    =1,\\
    \tilde{C}(\{o,o,o,o\},&\{\overline{1},\overline{1},\overline{1},\overline{1}\})
    =
    \tilde{C}(\{o,o,o,o\},\{\overline{2},\overline{2},\overline{2},\overline{2}\})
    =\sin^4\chi,\\
    \tilde{C}(\{e,o,e,o\},&\{\overline{1},\overline{2},\overline{1},\overline{2}\})
    =
    \tilde{C}(\{e,o,e,o\},\{\overline{2},\overline{1},\overline{2},\overline{1}\})\\
    =&
    \tilde{C}(\{o,e,o,e\},\{\overline{1},\overline{1},\overline{2},\overline{2}\})
    =
    \tilde{C}(\{o,e,o,e\},\{\overline{2},\overline{2},\overline{1},\overline{1}\})
    =\sin^2\chi,\\
    \tilde{C}(\{e,o,o,e\},&\{\overline{1},\overline{2},\overline{1},\overline{1}\})
    =
    \tilde{C}(\{e,o,o,e\},\{\overline{2},\overline{1},\overline{2},\overline{2}\})\\
    =&
    \tilde{C}(\{o,e,e,o\},\{\overline{1},\overline{1},\overline{2},\overline{1}\})
    =
    \tilde{C}(\{o,e,e,o\},\{\overline{2},\overline{2},\overline{1},\overline{2}\})
    =\sin^2\chi\left(1+\frac{1}{4}\cos^2\chi\right),\\
    \tilde{C}(\{o,o,e,e\},&\{\overline{1},\overline{1},\overline{1},\overline{2}\})
    =
    \tilde{C}(\{o,o,e,e\},\{\overline{2},\overline{2},\overline{2},\overline{1}\})\\
    =&
    \tilde{C}(\{e,e,o,o\},\{\overline{1},\overline{2},\overline{2},\overline{2}\})
    =
    \tilde{C}(\{e,e,o,o\},\{\overline{2},\overline{1},\overline{1},\overline{1}\})
    =\sin^2\chi\left(1-\frac{1}{4}\cos^2\chi\right),
\end{align*}
where $\overline{1}=(1,+)$ and $\overline{2}=(1,-)$.
The same contribution comes from the other Fermi surface $\lambda=2$.

We assume that the pair potential $\Delta$ consists only of the lowest LL.
Substituting the formulas derived in Ref.~\cite{adachi2003effects}
\begin{align}
    e^{i\rho\bm{v}\cdot\bm{\Pi}}
    \psi_{0q}(\bm{r})
    =&
    \frac{1}{\sqrt{\pi^{1/2}r_HL_y}}
    e^{-(|\beta|^2-\beta^2)/4}
    e^{-(1/2)(x/r_H+qr_H+\beta)^2+iqy},
    \tag{A25}\\
    e^{i\rho\bm{v}\cdot\bm{\Pi}^{*}}
    \psi^{*}_{0q}(\bm{r})
    =&
    \frac{1}{\sqrt{\pi^{1/2}r_HL_y}}
    e^{-(|\beta|^2-\beta^{*2})/4}
    e^{-(1/2)(x/r_H+qr_H-\beta^*)^2-iqy},
    \tag{A26}
\end{align}
where $\beta=\rho v_{\rm F}\hat{p}_{-}/r_H$, into Eq.~\eqref{eq:ap-14} and performing the integral with respect to $\bm{r}$, we obtain the expression,
\begin{equation}
    \mathcal{F}^{(4)}
    =
    \sqrt{\frac{\pi}{2}}
    \frac{TN(0)}{r_HL_y}
    \sum_{\{j_i\},\{q_i\}}
    \delta_{q_1+q_3,q_2+q_4}
    e^{-r_H^2(q_{13}^2+q_{24}^2)/4}
    \mathrm{Re}\left[V_4(\{j_i\},\{q_i\})
    \Delta^{*}_{j_1}(q_1)
    \Delta_{j_2}(q_2)
    \Delta^{*}_{j_3}(q_3)
    \Delta_{j_4}(q_4)\right],
    \tag{A27}
\end{equation}
with
\begin{align}
    V_4(\{j_i\},\{q_i\})
    =&
    \int^{\infty}_{0}\frac{d\rho_1d\rho_2d\rho_3}{\sinh[2\pi T(\rho_1+\rho_2+\rho_3)]}
    F(\{\rho_i\},\{j_i
    \})
    \left\langle I_4(\{\beta_i\},\{q_i\})|_{\beta_4=0}\right\rangle_{\hat{\bm{k}}},
    \tag{A28}\\
    F(\{\rho_i\},\{j_i\})
    =&\sum_{\{\overline{\nu}_i\}}
    \tilde{C}
    \cos(\rho_1\alpha_1+\rho_2\alpha_2+\rho_3\alpha_3),
    \tag{A29}\\
    \ln I_4(\{\beta_i\},\{q_i\})
    =&-\frac{1}{4}\sum_{i=1}^{4}|\beta_i|^2
    -\frac{1}{8}(\beta_{13}^{*2}+\beta_{24}^{2})
    -\frac{1}{4}(\beta_{1}^{*}+\beta_{3}^{*})(\beta_{2}+\beta_{4})
    +\frac{r_H}{2}(q_{13}\beta_{13}^{*}-q_{24}\beta_{24}),
    \tag{A30}
\end{align}
where $q_{ij}=q_i-q_j$, and $\beta_{ij}=\beta_{i}-\beta_{j}$.

\section{Appendix B: Maki parameter}
The upper critical field at zero temperature for $\Delta_e$-channel in the orbital (Pauli) limit  can be obtained as the solution of the equation
\begin{equation}
    E_{N}^{e}(H,T=0)=0,
    \tag{B1}
\end{equation}
with $h'=0\ (v_{\rm F}=0)$. The results are
\begin{align}
    H_{\rm c2}^{e,0,\mathrm{orb}}(T=0)
    =&\frac{2\pi^2}{C}\frac{(T_{\mathrm{c}0}^{e})^2}{ev_{\rm F}^{2}},
    \tag{B2}\\
    H_{\rm c2}^{e,0,P}(T=0)
    =&\frac{\pi}{C}\frac{mT_{\mathrm{c}0}^{e}}{e\cos\chi},
    \tag{B3}
\end{align}
where $\ln C=\gamma\simeq 0.5772$ is the Euler's constant.
Then, the Maki parameter~\cite{maki1966effect} is given by the ratio of $H_{c2}^{e,0,\mathrm{orb}}(T=0)$ to $H_{c2}^{e,0,P}(T=0)$,
\begin{align}
    \alpha_M
    =&\sqrt{2}
    \frac{H_{\rm c2}^{e,0,\mathrm{orb}}(T=0)}{H_{\rm c2}^{e,0,P}(T=0)}\notag\\
    =&2\sqrt{2}\pi\frac{T_{\mathrm{c}0}^{e}\cos\chi}{mv_{\rm F}^2}.
    \tag{B4}
\end{align}

\end{widetext}

\end{document}